\documentclass[aps,prd,onecolumn,nofootinbib,preprintnumbers,superscriptaddress]{revtex4}
\usepackage[mathscr]{eucal}
\usepackage{color}
\usepackage{graphicx}

\usepackage{epsf}
\usepackage{bm}

\usepackage{dcolumn}
\usepackage{bm}
\usepackage{amsmath}
\usepackage{amssymb}
\usepackage{amsfonts}
\usepackage{amsthm}
\usepackage{amscd}
\usepackage{graphicx}

\def\slashchar#1{\setbox0=\hbox{$#1$}           
   \dimen0=\wd0                                     
   \setbox1=\hbox{/} \dimen1=\wd1                   
   \ifdim\dimen0>\dimen1                            
      \rlap{\hbox to \dimen0{\hfil/\hfil}}          
      #1                                            
   \else                                            
      \rlap{\hbox to \dimen1{\hfil$#1$\hfil}}       
      /                                             
   \fi}

\def \sec{\begin{section}}
\def \esec{\end{section}}

\def \be{\begin{equation}}
\def \ee{\end{equation}}

\def \beq{\begin{equation}}
\def \eeq{\end{equation}}

\def \pr {\partial}

\begin{document}

\title{Anomaly matching condition in two-dimensional systems}

\author{O. Dubinkin}
\affiliation{Institute for Theoretical and Experimental Physics, B. Cheryomushkinskaya 25, Moscow 117218, Russia.}
\affiliation{Skolkovo Institute of Science and Technology, Skolkovo Innovation Center, Moscow 143026, Russia.}
\author{A. Gorsky}
\affiliation{Institute for Information Transmission Problems, B.Karetnyi 19, Moscow, Russia.}
\affiliation{Moscow Institute of Physics and Technology, Dolgoprudny 141700, Russia.}
\author{E. Gubankova}
\affiliation{Institute for Theoretical and Experimental Physics, B. Cheryomushkinskaya 25, Moscow 117218, Russia.}
\affiliation{Skolkovo Institute of Science and Technology, Skolkovo Innovation Center, Moscow 143026, Russia.}

\begin{abstract}

Based on Son-Yamamoto relation obtained for transverse part of triangle axial anomaly in ${\rm QCD}_4$, we derive its analog
in two-dimensional system.
It connects the transverse part of mixed vector-axial current two-point function with diagonal vector and axial
current two-point functions. Being fully non-perturbative, this relation may be regarded as anomaly matching for conductivities
or certain transport coefficients depending on the system. 
We consider the holographic RG flows in holographic
Yang-Mills-Chern-Simons theory via the Hamilton-Jacobi equation with respect to the radial coordinate.
Within this holographic model it is found that the RG flows for the following relations are diagonal: Son-Yamamoto relation and 
the left-right polarization operator. Thus the Son-Yamamoto relation holds at wide range of  energy scales.     
\\

Keywords: AdS/CFT correspondence, RG, two-point function, edge states
\end{abstract}

\maketitle

\section{Introduction}

\begin{figure}[!ht]
\includegraphics[width=0.15\textwidth]{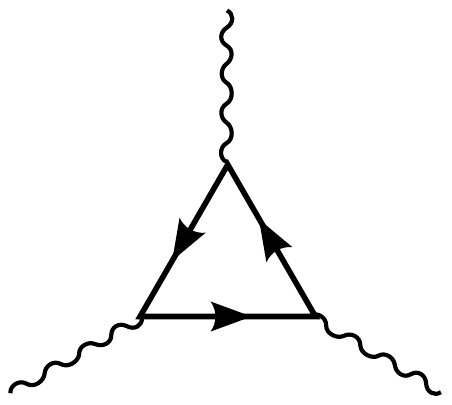}
\caption{Axial (ABJ) anomaly in (3+1)d: pion decay $\pi_0\rightarrow 2\gamma$.
Solid lines represent chiral fermions and wavy lines represent U(1) gauge bosons.}
\label{anomaly3}
\end{figure}

The usefulness of anomalies is partially related to the fact that they are exact and can be determined at strong coupling. This is a consequence
of certain non-renormalization properties and allows nonperturbative insight. Indeed ABJ axial anomaly can be captured perturbatively by one-loop Feynmann diagram. However, the result is non-perturbative, being exact from low to high energies since the anomaly
reflects the spectral flow at all scales.  Recently Son and Yamamoto derived an anomaly matching
condition which relates the $U(1)^3$ AVV triangle anomaly \cite{son-yamamoto}, Fig.(\ref{anomaly3}), to the two-point VV  and AA current functions, 
where V refers to vector current and A to the axial current. The result was obtained via holography and can be regarded as a non-perturbative exact relation between three- and two-point current functions. 
They used a five-dimensional Yang-Mills action of holographic dual of QCD and considered a holographic mechanism of chiral symmetry breaking via  the boundary conditions for the gauge fields in the infrared. This class of holographic theories incorporate
a bottom-up AdS/QCD inspired models and top-down Sakai-Sugimoto model.

In this paper, we consider holographic dual of (1+1)-dimensional systems 
given by a three-dimensional action and derive the analog of Son-Yamamoto relation. 
Like in holographic QCD dual action
can be considered as the worldvolume action on the probe flavor brane therefore
it involves the 3d Yang-Mills and Chern-Simons terms. The  duals of the
(1+1) dimensional systems in this approach have been considered in \cite{Yee-Zahed2011,jensen2011}. 
We will discuss the case of the single flavor in the boundary nonabelian gauge theory
at large N. The cigar geometry implies that like in 4+1 case we have to consider
left and right copies of the gauge group in the 3d bulk theory reflecting the
global  $U(1)_L \times U(1)_R$ symmetry at the boundary. The 2d QCD enjoys the chiral symmetry breaking 
\cite{zhitnitsky} via chiral condensate formation. There is  the pion-like
degree of freedom whose mass is related with the fermion mass via analog of    
the GOR relation. 

There is some important difference between the 4+1 and 2+1 bulk gauge theories. The 
Son-Yamamoto relation has been derived in 5-dim theory taking into account that the
contribution of CS terms is suppressed by the large t-Hooft coupling. Therefore it was
possible first consider the equation of motion without CS term derive the constant
Wronskian condition and then treat the CS term as a kind of perturbation. The situation
in 2+1 is different and there is no any suppression of CS term anymore  which 
is crucial for the imposing  the self-consistent boundary conditions \cite{Yee-Zahed2011}. 
Therefore we have to consider the equation of motion including the CS terms which 
have the opposite signs for the left and right fields. Therefore there is no naive
constant Wronskian condition and therefore naive 1+1 analog of Son-Yamamoto relation.
However the numerical analysis of the equations of motion demonstrates that the Wronskian 
exhibits a platoe in the very wide interval of the radial holographic coordinate and the
transition to the platoe is very sharp. One could also have in mind 
the formal regime when YM terms dominate. Therefore we can explore the constant Wronskian
condition with some reservations in 2+1 case as well.


It was shown in \cite{son-yamamoto} that the SY relation is consistent with Vainshtein relation \cite{Vainshtein2002}
for the magnetic susceptibility of the quark condensate in QCD introduced in \cite{Ioffe-Smilga1984}.
To some extend it can be considered as a way of its derivation. However, in two dimensions the OPE for vector-axial correlator 
trivially reduces the four-fermion operator to the square of the chiral condensate due to the $2$D chiral algebra.
As a result we obtain from the Son-Yamamoto relation an estimate for the pion decay constant.
We note that it is derived in the
region when the application of the low-energy theory is questionable. Hence this result 
should be taken with some reservation and deserves the additional study.

An additional question concerns RG flow of our holographic model. This question is related to renormalization
and regularization of effective theories in holography, that was solved along two avenues. First is the method of standard holographic renormalization that
involves the cancellation of all cut-off related divergences from the gravity on-shell action by adding the counterterms on the cut-off boundary surface
and the subsequent removal of cut-off \cite{review}. Holographic renormalization has been used in calculation of two-point functions in deformed CFT \cite{review}. In parallel development, the Hamilton-Jacobi equation was advocated to use for renormalization in order to separate terms in the bulk
on-shell action, which can be written as local functions of boundary data. The remaining non-local expression was identified, according
to the AdS/CFT prescription, with the generating functional of a boundary field theory \cite{verlinde-verlinde,martelli-mueck}. In Hamilton-Jacobi equation the bulk radial coordinate is treated as the time variable, which is consistent with holographic identification of radial coordinate with RG energy scale.
The second approach provides correct results for anomalies  and gives a simple
description of RF flow in deformed CFT's \cite{papadimitriou}. We apply  Hamilton-Jacobi equation in the bulk theory to the Yang-Mills-Chern-Simons holographic action similar to \cite{dubinkin-gorsky-milehin2015} and demonstrate that SY relation is diagonal with respect to 
holographic RG flow.      
\\
\\
\\

\begin{figure}[!ht]
\includegraphics[width=0.2\textwidth]{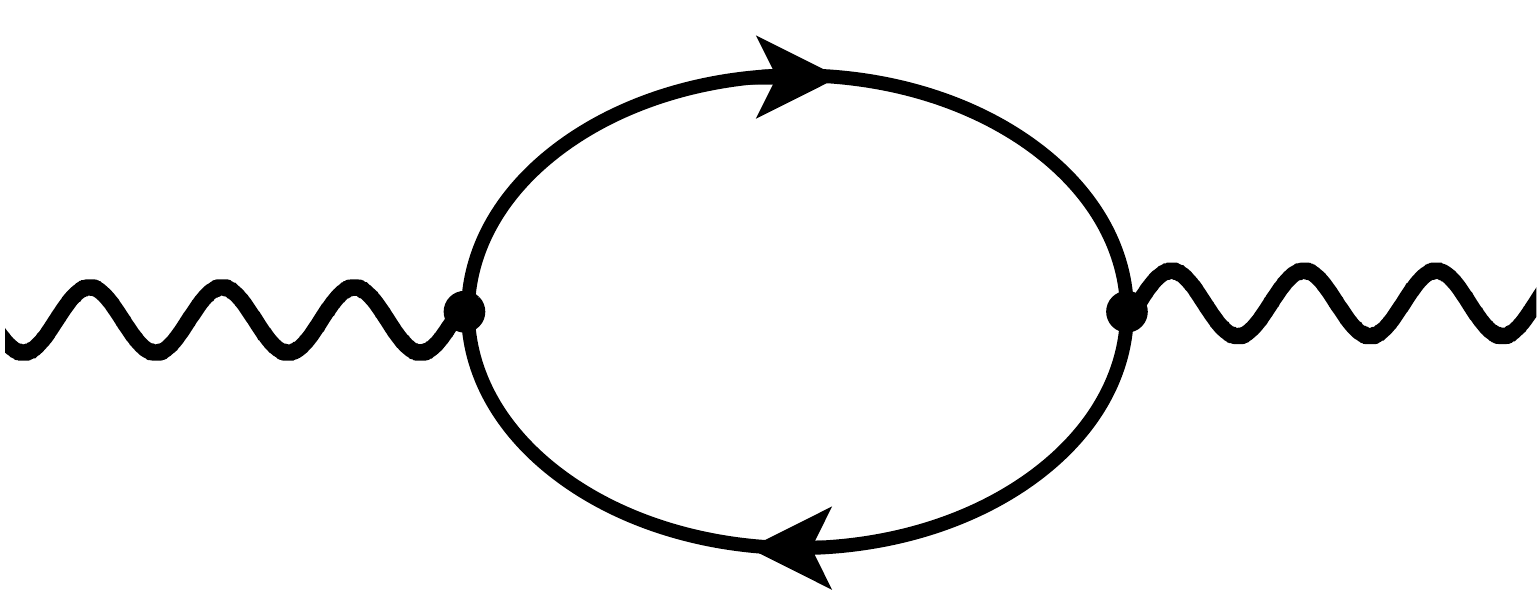}
\caption{Parity violating anomaly in (1+1)d: mass generation $m\bar{\psi}\psi$.
Solid lines represent chiral fermions and wavy lines represent U(1) gauge bosons.}
\label{anomaly1}
\end{figure}

The paper is organized as follows. We derive the two-dimensional Son-Yamamoto relation in the Section 2. In Section 3 we check the Son-Yamamoto relation
in the small and large $Q^2$ limits, and obtain an estimate for the pion decay constant.
In  Section 4 we demonstrate using the Hamilton-Jacobi  equations in the bulk theory that Son-Yamamoto relation
is diagonal under the RG flows. Section 5 is devoted to the comparison of
our results for the $1+1$-dimensional Son-Yamamoto relation with the one obtained in the $3+1$-dimensional QCD.
The results are summarized in Conclusion and technical details are collected in Appendixes A and B.

\section{Model and Son-Yamamoto relation}\label{section:1}

We consider chiral dynamics in two dimensions. Chiral symmetry is $U(1)_L\times U(1)_R$, that corresponds to the conserved left- and right-handed currents.
According to AdS/CFT duality, there are left $A_L$ and right-handed $A_R$ gauge fields in a three-dimensional  dual model.  
The $3$D dual action involves three-dimensional Maxwell theory and the topological Chern-Simons term
\begin{equation}
S=S_{M}+S_{CS}= S_{M}(A_L) +S_{M}(A_R) + S_{CS}(A_L)-S_{SC}(A_R)
\label{action}
\end{equation}
where $S_{M}$ and $S_{CS}$ are defined as
\begin{equation}
S_{M}({\mathcal A})=\int d^2 x dz\ \left( f(z){\mathcal F}_{z\mu}^2 - \frac{1}{2g(z)}{\mathcal F}_{\mu\nu}^2 \right)
\label{yang-mills-action}
\end{equation}
and \cite{Yee-Zahed2011,Dunne1998}
\begin{equation}
S_{CS}({\mathcal A})=\kappa\int d^2x dz\ \left( {\mathcal A}\ast {\mathcal F}\right)
\label{chern-simons-action}
\end{equation}
with \cite{Yee-Zahed2011,Dunne1998}
\begin{equation}
\kappa= \frac{N_c}{4\pi},
\label{kappa}
\end{equation} 
and the dual field strength is $\ast {\mathcal F}_{\mu}=\frac{1}{2}\varepsilon_{\mu\nu\lambda}{\mathcal F}^{\nu\lambda}$.

The IR brane is located at $z=0$ and the UV boundary of the asymptotic ${\rm AdS}_3$ space is located at $z=z_0$.
It is convenient to use vector $V$ and axial $A$ gauge fields
\begin{equation}
A_L=V +A,\quad A_R=V-A
\label{va}
\end{equation}
which obey Neumann and Dirichlet boundary conditions in the IR, respectively,
\begin{equation}
{\rm IR:}\;\;\partial_zV_{\mu}(z=0)=0,\quad A_{\mu}(z=0)=0
\label{neumann-dirichlet}
\end{equation}
and $V(-z)=V(z)$ is parity even, $A(-z)=-A(z)$ is parity odd. Making use of the decomposition eq.(\ref{va}), the Maxwell and Chern-Simons terms in
the action eq.(\ref{action}), are given by
\begin{eqnarray}
S_{M} &=& \int d^2xdz\ \left(f(z)^2(F_{Vz\mu}^2+F_{Az\mu}^2)
-\frac{1}{2g(z)^2}(F_{V\mu\nu}^2 + F_{A\mu\nu}^2)\right) \label{yang-mills2}\\
S_{CS} &=& 2\kappa \int d^2xdz \left(V_{\mu}\ast F_{A\mu}+ A_{\mu}*F_{V\mu}\right)
\label{chern-simons2}
\end{eqnarray}
We will work in the radial gauge: $V_z=A_z=0$ and assume there is a translation invariance along the boundary "UV"-brane, and perform the Fourier transform for gauge fields:
\begin{equation}
V_{\mu}(x,z) = \int \frac{d^2q}{(2\pi)^2} {\rm e}^{-iqx}V(q,z)
\label{}
\end{equation}
and the same for the axial field $A_{\mu}$. 
Substituting these expressions into the action, we can write down the holographic Maxwell and Chern-Simons terms in $3$D explicitly
\begin{eqnarray}
S_{M} &=& \int \frac{d^2q}{(2\pi)^2}dz\ \left(f(z)^2((\partial_z V_{\mu})^2+(\partial_z A_{\mu})^2)
-\frac{1}{2g(z)^2}(F_{V\mu\nu}^2 + F_{A\mu\nu}^2)\right), \label{yang-mills}\\
S_{CS} &=& 2\kappa \int \frac{d^2q}{(2\pi)^2}dz\ \varepsilon^{\mu\nu}\left(\partial_zA_{\mu}V_{\nu}+\partial_zV_{\mu}A_{\nu}\right),
\label{chern-simons}
\end{eqnarray}
where we used the convention \cite{Mottola2014} $\varepsilon^{z\mu\nu}\equiv \varepsilon^{\mu\nu}$. Here
$\varepsilon_{\mu\nu}=-\varepsilon_{\nu\mu}$ is the two-dimensional anti-symmetric symbol, $\varepsilon_{01}=1=-\varepsilon^{01}$ which obeys
\begin{equation}
\varepsilon^{\mu\lambda}\varepsilon_{\nu\rho}=-\delta^{\mu}_{\nu}\delta^{\lambda}_{\rho}+\delta^{\mu}_{\rho}\delta^{\lambda}_{\nu},\;\;
\varepsilon^{\mu\nu}\varepsilon_{\nu\rho}=\delta^{\mu}_{\rho}.
\label{epsilon}
\end{equation}
Following Son and Yamamoto \cite{son-yamamoto} we are interested in the transversal part of the correlators. Further for short we will omit perpendicular projectors $P^{\perp}_{\mu\nu}=\eta_{\mu\nu}-\frac{q_{\mu}q_{\nu}}{q^2}$ in the expressions for gauge fields $V$ and $A$. 
However, it can be easily reinstated in the resulting formulas by substitute $\delta_{\mu\nu}\rightarrow P^{\perp}_{\mu\nu}$.
We perform the following field decomposition:
\begin{equation}
V_{\mu}(q)=V(q,z)V_{0\mu}(q),\quad A_{\mu}(q)=A(q,z)A_{0\mu},
\label{}
\end{equation}
where $V_{0\mu}$, $A_{0\mu}$ are the sources of the corresponding boundary currents. 
We require the Dirichlet boundary conditions in the UV
\begin{equation}
{\rm UV:}\;\; V(q,z)|^{z=z_0}=1,\;\; A(q,z)|^{z=z_0}=1,
\label{condition}
\end{equation}
therefore sources coincide with the bulk gauge fields at the boundary.
According to the AdS/CFT prescription to obtain correlation functions for currents, we need to vary the action evaluated on the classical solutions with respect to the corresponding sources of the currents, $V_{0\mu}$ and $A_{0\mu}$. 

Next let us remind of the observation made by Son and Yamamoto, that
made possible to derive the relation between three- and two-point functions \cite{son-yamamoto}. In our case we will obtain the relation between diagonal
and mixed two-point functions. 
The linearized equation of motion in the pure Maxwell theory is
\begin{eqnarray}
\partial_z\left(f^2\partial_z V\right) - \partial_{\mu}\left(\frac{1}{g^2}\partial_{\mu} V\right)=0,
\label{}
\end{eqnarray}
which due to the translation invariance along the boundary direction is
\begin{eqnarray}
\partial_z\left( f^2(z)\partial_z V(q,z)\right)+\frac{q^2}{g^2(z)}V(q,z) = 0.
\label{solutions}
\end{eqnarray} 
Here we did not take into account the Chern-Simons term. 
The same equation of motion is satisfied by the axial gauge field, i.e. we have 
\begin{eqnarray}
V^{\prime\prime}+\frac{\partial f^2}{f^2}V^{\prime} +\frac{q^2}{g^2}V &=& 0\\
A^{\prime\prime}+\frac{\partial f^2}{f^2}A^{\prime} +\frac{q^2}{g^2}A &=& 0
\label{solutions}
\end{eqnarray}
Since $V$ and $A$ are linearly independent solutions of the same equation, we have \cite{son-yamamoto}
\begin{equation}
V(q,z)\pr A(q,z) -A(q,z)\pr V(q,z) = W(q){\rm e}^{-\int \partial f^2/f^2}
\label{}
\end{equation}
which can be written as
\begin{equation}
f^2(z)\big( V(q,z)\pr_zA(q,z) -A(q,z)\pr_zV(q,z)\big) = W(q)
\label{wronskian}
\end{equation}
where $W(q)$ is $z$-independent Wronskian. The independence on $z$ of the combination in eq.(\ref{wronskian}) is crucial
to obtain Son-Yamamoto relation and it is responsible for the unique properties of the RG equations for the diagonal and mixed correlators. 
Relation for Wronskian eq.(\ref{wronskian}) is obtained in the pure Maxwell theory. 

In Appendix B we include the Chern-Simons term and regulate the Maxwell-Chern-Simons theory using dimensional regularization. We show numerically that
the logarithmic divergences characteristic of the $(1+1)$ dimensional boundary theory \cite{Yee-Zahed2011} are regularized: solutions for the gauge fields $V(Q,z)$ and $A(Q,z)$ converge in the UV. It justifies the use of the finite boundary conditions in eq.(\ref{condition}). Furthermore, Wronskian for the regulated Maxwell-Chern-Simons theory has a platou starting from some radial coordinate $z>z_p$.   
Therefore it is legitimate to take $z$-independent Wronskian for large $Q^2$ or when the
coefficient in front of the Chern-Simons term is small.   
\\
\\
\\
{\bf Diagonal correlators}

Now we are ready to obtain two-point diagonal correlators for the vector $\langle V_{\mu}V_{\nu}\rangle$ and axial $\langle A_{\mu}A_{\nu}\rangle$
fields in $(1+1)$ dimensions. Varying the action twice with respect to the boundary value $V_{0\mu}$, we obtain the vector current two-point correlator   
\begin{eqnarray}
&& \langle j_{\mu}(q)j_{\nu}(-q)\rangle = \int d^2x\  {\rm e}^{iqx} \langle j_{\mu}(x)j_{\nu}(0)\rangle  \\
&& \frac{\delta^2S_{YM}}{\delta V_{0\mu}(q)\delta V_{0\nu}(-q)} = \langle j_{\mu}(q)j_{\nu}(-q)\rangle
\label{two-point}
\end{eqnarray}
where the Yang-Mills action is given by eq.(\ref{yang-mills}). Intregrating the first term of eq.(\ref{yang-mills}) by parts we obtain
\begin{equation}
S_{YM} = 2\left.\int \frac{d^2q}{(2\pi)^2}f^2(z){\mathcal A}_{\mu}(q,z)\partial_z{\mathcal A}^{\mu}(q,z)\right|^{z=z_0}_0
\label{yang-mills2}
\end{equation}
where ${\mathcal A}$ stands either for $V$ or $A$; and
the action is evaluated at the solutions eq.(\ref{solutions}). 
Varying eq.(\ref{yang-mills2}) twice with respect to the boundary values $V_0$ eq.(\ref{two-point}) 
\begin{equation}
\langle j_{\mu}j_{\nu}\rangle= \left.2f^2(z)V(q,z)V^{\prime}(q,z)\right|^{z=z_0}\delta_{\mu\nu}
\label{}
\end{equation}
where we introduced the notation $V'=\pr_zV$. Substituting the boundary conditions eq.(\ref{condition}), we obtain for the two-point correlation functions
\begin{eqnarray}
&& \langle j_{\mu}j_{\nu}\rangle= \left.2f^2V^{\prime}\right|^{z=z_0}\delta_{\mu\nu} \equiv \Pi_{V}(q)\delta_{\mu\nu},\\
\label{vv}
&& \langle j^5_{\mu}j^5_{\nu}\rangle= \left.2f^2A^{\prime}\right|^{z=z_0}\delta_{\mu\nu} \equiv \Pi_{A}(q)\delta_{\mu\nu},
\label{aa}
\end{eqnarray}
where we introduced (dimensionless) polarization operators $\Pi_V$ and $\Pi_A$. Therefore the polarization operators are given by
\begin{equation}
\Pi_V=\left.2f^2V^{\prime}\right|^{z=z_0},\,\,\Pi_A=\left.2f^2A^{\prime}\right|^{z=z_0}.
\label{polarization-operators}
\end{equation}
Eq.(\ref{polarization-operators}) represents the known expression for a diagonal conductivity obtained from Kubo formula.
From eq.(\ref{yang-mills2}), the diagonal current-current correlator is given by 
\begin{equation}
\langle j_{i}j_{j}\rangle=\left.2f^2\frac{\partial_z {\mathcal A}_{i}}{{\mathcal A}_{i}}\right|^{z=z_0}\delta_{ij}
\label{}
\end{equation}
where via Kubo formula the relation to the conductivity $\sigma_{ij}$ is $\langle j_{i}j_{j}\rangle=\sigma_{ij}$.

Using the expression for the Wronskian eq.(\ref{wronskian}) and the boundary conditions eq.(\ref{condition}), we obtain the following relation
\begin{equation}
f^2\left(VA^{\prime}-AV^{\prime}\right)= \frac{1}{2}\left(\Pi_A-\Pi_V\right) =W(q),
\label{pia-piv}
\end{equation}
which we use further.
Since combination in eq.(\ref{wronskian}) does not depend on $z$, it can be estimated at any point, for example at $z=z_0$
where the polarization operators are defined by eq.(\ref{polarization-operators}).     
Wronskian equation (\ref{pia-piv}) is the crucial formula to establish a relation between diagonal and mixed current correlators.
\\
\\
{\bf Mixed correlator}

Now our aim is to obtain mixed correlator for axial and vector fields $\langle V_\mu A_\nu\rangle$. Easy to see that the only contribution to this correlator function is coming from Chern-Simons term eq.(\ref{chern-simons}). After fourier transformation Chern-Simons term (\ref{chern-simons}) can be rewritten as
\begin{equation}
S_{CS} = 2\kappa\int\frac{d^2q}{(2\pi)^2}dz\varepsilon^{\rho\sigma}\left(\partial_zA_{\rho}(-q)V_{\sigma}(q)
-\partial_zV_{\sigma}(q)A_{\rho}(-q)\right).
\label{chern-simons100}
\end{equation}
As in \cite{son-yamamoto,Gorsky-Krikun2009}, we can add the surface term to the action,  
\begin{equation}
\delta S_{CS} = -2\kappa\int\frac{d^2q}{(2\pi)^2}dz\varepsilon^{\rho\sigma}\partial_z\left(A_{\rho}(-q)V_{\sigma}(q)\right)
\label{chern-simons200}
\end{equation}
that is equivalent to the gauge transformation done
in \cite{son-yamamoto}. The Chern-Simons term becomes $S_{CS}+\delta S_{CS}\rightarrow S_{CS}$,
\begin{equation}
S_{CS} = -4\kappa\int\frac{d^2q}{(2\pi)^2}dz\varepsilon^{\rho\sigma}A_{\rho}(-q)\partial_zV_{\sigma}(q)
\label{chern-simons-term}
\end{equation}
Varying twice the (new) Chern-Simons term with respect to the boundary fields 
\begin{equation}
\frac{\delta^2S_{CS}}{\delta V_{0\mu}\delta A_{0\nu}} =\langle j_{\mu}j^5_{\nu}\rangle
\label{}
\end{equation}
we obtain
\begin{equation}
\frac{\delta^2S_{CS}}{\delta V_{0\mu}\delta A_{0\nu}}=\langle j_{\mu}j^5_{\nu}\rangle = -2\kappa \varepsilon^{\nu\mu}\left.AV\right|^{z_0}
+2\kappa \varepsilon^{\nu\mu}\int^{z_0} 
dz(A^{\prime}V-V^{\prime}A)\equiv \frac{1}{2\pi}\varepsilon^{\mu\nu}w_T(q),
\label{mixed}
\end{equation}
where we introduced a (dimensional) transversal part of the vector-axial current correlator $w_T$.
Therefore we have
\begin{equation}
w_T=4\pi\kappa-4\pi\kappa\int^{z_0}dz (A^{\prime}V-V^{\prime}A).
\label{}
\end{equation}
From eq.(\ref{pia-piv}) it can be written as
\begin{equation}
w_T(Q^2) = 4\pi\kappa-4\pi\kappa \int_0^{z_0} \frac{dz}{2f^2(z)}(\Pi_A(Q^2)-\Pi_V(Q^2)),
\label{son-yamamoto1}
\end{equation}
where we used and $Q^2=-q^2$. In the context of the ${\rm QCD}_2$, $4\pi\kappa=N_c$ eq.(\ref{kappa}).
Relation between the mixed and diagonal correlators eq.(\ref{son-yamamoto1}) is the $(1+1)$-dimensional analog of the Son-Yamamoto relation, which was originally obtained
for the QCD in $(3+1)$-dimensions \cite{son-yamamoto}.
One can think of the Son-Yamamoto relation eq.(\ref{son-yamamoto1}) as the expansion in the large $Q^2$.
The first term (normalized by eq.(\ref{mixed}) to be equal to $4\pi\kappa=N_c$) is the perturbative contribution of the axial anomaly, which is obtained from PT loop calculation. 
In $QCD$, the integral of the metric factor $\int 1/f^2$ produces $1/f_{\pi}^2$ where $f_{\pi}$ is the pion
decay constant \cite{dubinkin-gorsky-milehin2015}. 

We rewrite the Son-Yamamoto relation in the form
\begin{equation}
(SY) = w_T-4\pi\kappa+ 4\pi\kappa\int^{z_0} \frac{dz}{2f^2(z)}(\Pi_A-\Pi_V) =0.
\label{son-yamamoto-relation}
\end{equation}
This relation holds for any metric factors $f(z)$ and $g(z)$.

Generally, we decompose the axial anomaly as
\begin{equation}
\langle j_{\mu}j_{\nu}^5 \rangle= \frac{1}{2\pi}P_{\mu}^{\alpha\,\perp}\left(P_{\nu}^{\beta\,\perp}w_T+P_{\nu}^{\beta\,\parallel}w_L
\right)\varepsilon_{\alpha\beta},
\label{decomposition}
\end{equation} 
where the transverse and longitudinal projection ternsors are $P_{\mu}^{\alpha\perp}=\eta_{\mu}^{\alpha}-q_{\mu}q^{\alpha}/q^2$ and 
$P_{\mu}^{\alpha\parallel}=q_{\mu}q^{\alpha}/q^2$, respectively. 
In eq.(\ref{decomposition}), the perturbative $w\,$s are given by \cite{Mottola2014}
\begin{equation}
w_T=4\pi\kappa=N_c,\;\;w_L=8\pi\kappa=2N_c
\label{wT-wL}
\end{equation} 
The same relation $w_L=2w_T$ holds in the  ${\rm QCD}_4$.
 
We give another represenation for the Son-Yamamoto relation through the left-right correlator. 
The left-right correlator $\langle LR \rangle$, which is the measure of the chiral symmetry breaking, can be expressed through the diagonal correlators
$\langle VV\rangle$ and $\langle AA\rangle$ as
\begin{equation}
\Pi_{LR} = \Pi_A-\Pi_V.
\label{pi_lr}
\end{equation}
Using the definition of $w_T$ eq.(\ref{decomposition}),
we rewrite the Son-Yamamoto relation eq.(\ref{son-yamamoto1}) as
\begin{equation}
\langle j^{L}_{\mu}j^{R}_{\nu} \rangle^{\perp}=P_{\mu}^{\alpha\,\perp}P_{\nu}^{\beta\,\perp}\varepsilon_{\alpha\beta}
\left(4\kappa - 4\kappa\int_0^{z_0}\frac{dz}{2f^2}\Pi_{LR}\right)
\label{son-yamamoto-relation2}
\end{equation}
where $j_{\mu}^L=\bar{\psi}_L\gamma_{\mu}\psi_L$ is the left-handed current and $j_{\mu}^R=\bar{\psi}_R\gamma_{\mu}\psi_R$ is the right-handed current.
Equation (\ref{son-yamamoto-relation2}) agrees with the one obtained in the Schwinger model in \cite{Mottola2014}. The proportionallity coefficient
also agrees with the two-dimensional perturbative calculations.  
In what follows we use the both representations of the Son-Yamamoto relation, eq.(\ref{son-yamamoto-relation}) and eq.(\ref{son-yamamoto-relation2}).

\section{Checking the Son-Yamamoto relation. The pion decay constant}\label{section:4}

Let us check if the Son-Yamamoto relation eq.(\ref{son-yamamoto-relation}) is satisfied in a model independent setting. In order to estimate the individual
two-point current correlators we will consider the two opposite limits of small and large momenta $Q^2$, where some simplifications can be done.
Also we will make an estimate for the decay constant.
\\
\\
{\bf Regime of small $Q^2$}

First, we consider the limit of small $Q^2\ll \Lambda^2$. In this case we estimate the Son-Yamamoto relation at the point $z_0\rightarrow 0$. In the next section \ref{section:2}, we associate the UV boundary cutoff $z_0$
with the RG scale in the Hamilton-Jacobi equation. Therefore the limit when the UV cuoff is taken to be zero corresponds to the field theory in the regime of the low energy/momentum. 
In the limit $z_0 = 0$, the different boundary conditions eq.(\ref{neumann-dirichlet})
enable us to simplify the holographic action. 
As discussed in \cite{dubinkin-gorsky-milehin2015}, in the Yang-Mills action, we can neglect $\partial_z V_{\mu}=0$, however we approximate
$\partial_z A_{\mu} = \frac{A_{\mu}}{z_0}$. Therefore we can write to the leading order
\begin{eqnarray}
S_{YM} = z_0\int d^2x\left(\frac{f^2}{z_0^2}A_{\mu}^2-\frac{1}{2g^2}F_{V\mu\nu}^2\right)
\label{}
\end{eqnarray}  
that gives
\begin{eqnarray}
\Pi_A = \frac{2f^2(z_0)}{z_0},\quad \Pi_V = 0
\label{}
\end{eqnarray} 
and together with the intergral 
\begin{eqnarray}
\int\frac{dz}{2f^2}\left(\Pi_A-\Pi_V\right)= \frac{z_0}{2f^2(z_0)}\frac{2f^2(z_0)}{z_0}=1
\label{limit1}
\end{eqnarray} 
In the Chern-Simons action eq.(\ref{chern-simons2},\ref{chern-simons100}) and the boundary term eq.(\ref{chern-simons200}), we can neglect the term $\varepsilon^{\mu\nu\lambda}A_{\mu}F_{V\nu\lambda}$, but again
take $\partial_z A_{\mu}= \frac{A_{\mu}}{z_0}$ in the term $\varepsilon^{\mu\nu\lambda}V_{\mu}F_{A\nu\lambda}$. 
Approximating the integral over $z$, we have to the leading order
\begin{eqnarray}
S_{CS} &=& 2\kappa \int d^2x \varepsilon^{\mu\nu}A_{\mu}V_{\nu},\nonumber\\
\delta S_{CS} &=& -2\kappa \int d^2x \varepsilon^{\mu\nu}A_{\mu}V_{\nu} 
\label{}
\end{eqnarray} 
thus from eq.(\ref{mixed}) 
\begin{eqnarray}
w_T = 0
\label{limit2}
\end{eqnarray}
Combining together eqs.(\ref{limit1},\ref{limit2}), the Son-Yamamoto relation eq.(\ref{son-yamamoto-relation}) is satisfied at $z_0=0$, i.e.
it holds for small $Q^2$.
\\
\\
{\bf Regime of large $Q^2$}

Next, we consider the opposite limit $Q^2\gg \Lambda^2$, where we can use the operator product expansion. From
the Son-Yamamoto relation we will make an estimate for the decay constant.

In two dimensions the  fermion field has dimension $[\psi]=\sqrt{E}$, the gauge field $[F]=E$, $[A]=1$, the coupling of fermion with gauge fields $[g]=E$  
and the decay constant is dimensionless $[f_{\pi}]=1$. The anomalous divergence of the axial current is
\begin{equation}
\partial_{\mu}j_5^{\mu}=4\kappa g\tilde{F},
\label{axial-anomaly}
\end{equation} 
where the dual field strength equals to the $2$D electric field $\tilde{F}=\frac{1}{2}\varepsilon_{\mu\nu}F^{\mu\nu}=E$ which is the pseudoscalar. 
The Dirac matrices in the $2\times 2$ chiral representation are given by the Pauli matrices \cite{Mottola2014,Basar-Dunne2012}
\begin{equation}
\gamma_0=\sigma_1,
\,\,\gamma_1=-i\sigma_2,
\,\,\gamma_5=\gamma_0\gamma_1=\sigma_3,
\label{Dirac-matrices}
\end{equation}
where $\sigma_i\sigma_j=\delta_{ij}+i\epsilon_{ijk}\sigma_k$ with $i,j,k=1,2,3$. 
A special property of the gamma matrices in two dimensions is \cite{Mottola2014}
\begin{equation}
\gamma^{\mu}\gamma_5=\gamma_{\nu}\varepsilon^{\nu\mu},
\end{equation}
that gives for the spin-operator $\sigma_{\mu\nu}=\frac{1}{2i}(\gamma_{\mu}\gamma_{\nu}-\gamma_{\nu}\gamma_{\mu})$
\begin{equation}
\sigma_{\mu\nu}=i\gamma_5\varepsilon_{\mu\nu}.
\label{simplify}
\end{equation}
This property enables us to make significant simplifications in the diagrams of the OPE in the two dimensions, that follows next. 

We check the Son-Yamamoto relation at large virtualities and obtain the result for the decay constant.   
As in \cite{son-yamamoto}, we compare the OPE and the Son-Yamamoto relation for the two-point left-right current correlator.
Diagrams contributing to the $\langle j_{\mu}j_{\nu}^5\rangle$ in the OPE are the fermion loops, which are opened on two sides, have insertions of 
the (chiral) scalar $\langle\bar{\psi}\psi\rangle$ and
the spin-chiral $\langle\bar{\psi}\sigma_{\mu\nu}i\gamma_5\psi\rangle$ condensates with spin operator $\sigma_{\mu\nu}$, and different arrangements of a photon line 
in the fermion loop are possible .   
Therefore on one hand, the OPE is written as
\begin{equation}
\langle j_{\mu}^L j_{\nu}^{R}\rangle =\frac{1}{2}\langle j_{\mu}j_{\nu}^5 \rangle =P_{\mu}^{\alpha\,\perp}P_{\nu}^{\beta\,\perp}
\left(4\kappa\varepsilon_{\alpha\beta}+\frac{2g^2}{Q^2}O_{\alpha\beta}\right),
\end{equation} 
where the operator 
\begin{equation}
O_{\alpha\beta}=\frac{\langle (\bar{\psi}\gamma_{\alpha}\gamma_5\psi)(\bar{\psi}\gamma_{\beta}\psi)\rangle}{Q^2}, 
\end{equation}
is the four-fermion operator. Using the Fierz transformation and in the large-$N_c$ limit where the four-fermion operator factorizes we have
in the leading order
\begin{equation}
O_{\alpha\beta}=\frac{\langle\bar{\psi}\psi\rangle \langle\bar{\psi}\sigma_{\alpha\beta}i\gamma_5\psi\rangle}{Q^2}
= -\frac{\langle\bar{\psi}\psi\rangle^2}{Q^2},
\end{equation}
where we simplified the spin-chiral condensate with the help of eq.(\ref{simplify}).
The leading order OPE for the left-right current correlator is given by the operator dimension two
\begin{equation}
\langle j_{\mu}^L j_{\nu}^{R}\rangle^{\perp} =P_{\mu}^{\alpha\,\perp}P_{\nu}^{\beta\,\perp}\varepsilon_{\alpha\beta}
\left(4\kappa-\frac{2g^2}{Q^4}\langle\bar{\psi}\psi\rangle^2\right).
\label{left-right-correlator1}
\end{equation} 
On the other hand, the Son-Yamamoto relation is given by eq.(\ref{son-yamamoto-relation2})
\begin{equation}
\langle j^{L}_{\mu}j^{R}_{\nu} \rangle^{\perp}=P_{\mu}^{\alpha\,\perp}P_{\nu}^{\beta\,\perp}\varepsilon_{\alpha\beta}
\left(4\kappa - 4\kappa\int_0^{z_0}\frac{dz}{2f^2}\Pi_{LR}\right)
\label{LR1}
\end{equation}
The leading term in the OPE for $\Pi_{LR}$ is dimension two operator \cite{Shifman-Vainstein-Zakharov} 
\begin{equation}
\Pi_{LR} = -\frac{g^2}{Q^2}\frac{\langle(\bar{\psi}_L\gamma_{\mu}\psi_L)(\bar{\psi}_R\gamma_{\mu}\psi_R)\rangle}{Q^2}
=\frac{2g^2}{Q^4}\langle\bar{\psi}\psi \rangle^2.
\label{}
\end{equation}
Comparing terms proportional to $\langle\bar{\psi}\psi\rangle^2$ in eqs.(\ref{left-right-correlator1},\ref{LR1}), we find  
\begin{equation}
\frac{4\kappa}{f_{\pi}^2}= 1,
\label{equality}
\end{equation}
where we made the following identification of the integral with the decay constant \cite{son-yamamoto}
\begin{equation}
\frac{1}{f_{\pi}^2}=\int^{z_0}\frac{dz}{2f^2(z)}.
\end{equation}
Equation (\ref{equality}) relies completely on the Son-Yamamoto relation.
We consider it not as an exact result, but as an estimate for the $f_{\pi}$, because as discussed for ${\rm QCD}_4$ in \cite{son-yamamoto}
the Son-Yamamoto equation does not provide complete match for resonances at large virtualities. 
In the ${\rm QCD}_2$, the Chern-Simons $\kappa$ is proportinal to $N_c$ eq.(\ref{kappa}). Therefore we have from eq.(\ref{equality})
\begin{equation}
f_{\pi}^2\sim N_c,
\end{equation}
that agrees with estimate done in the weak coupling regime of 't Hooft solution 
$N_c\rightarrow\infty$ and $g^2N_c={\rm const}$ by Zhitnitsky in \cite{zhitnitsky}.
\\
\\
Also we can check the Son-Yamamoto relation using the parallel component.
The OPE for the parallel component is given by the diagram Fig.1 in \cite{zhitnitsky} and includes operator of dimension two
\begin{equation}
\langle j_{\mu}j_{\nu}^5\rangle^{\parallel}=P_{\mu}^{\alpha\,\perp}P_{\nu}^{\beta\,\parallel}\left(
4\kappa\varepsilon_{\alpha\beta}-\frac{2m\langle \bar{\psi}\sigma_{\alpha\beta}i\gamma_5\psi\rangle}{Q^2}
\right)=P_{\mu}^{\alpha\,\perp}P_{\nu}^{\beta\,\parallel}\varepsilon_{\alpha\beta}\left(4\kappa
+\frac{2m\langle \bar{\psi}\psi\rangle}{Q^2}\right).
\label{paral1}
\end{equation}
On the other hand, expanding the pion pole propagator \cite{zhitnitsky,Mottola2014}, we have
\begin{equation}
Q^2\frac{f_{\pi}^2}{Q^2+m_{\pi}^2}=f_{\pi}^2\left(1-\frac{m_{\pi}^2}{Q^2}\right).
\label{paral2}
\end{equation}
Comparing eqs.(\ref{paral1},\ref{paral2}), we obtain
\begin{equation}
f_{\pi}^2m_{\pi}^2=-2m\langle\bar{\psi}\psi\rangle,
\label{}
\end{equation}
which is the Gell-Mann-Oakes-Renner relation, i.e. we trivially satisfy the Son-Yamamoto relation.
Our calculations for the parallel component
rely on the pion pole dominance -- saturation of the two-point correlators by the pion pole contribution -- valid at small $Q^2$ \cite{Yee-Zahed2011}. 
Here we analytically continue it to the large $Q^2$.

\section{Hamilton-Jacobi equation}\label{section:2}

As was argued in \cite{verlinde-verlinde}, the holographic renormalization group equation can be obtained as a Hamiltonian-Jacobi equation
when time is identified with the radial $z$ coordinate. According to the AdS/CFT prescription it is written as 
\begin{equation}
\frac{\partial S}{\partial z_0} + H({\mathcal \pi}_{\alpha},{\mathcal A}_{\alpha},z_0)=0
\label{hamilton-jacobi}
\end{equation}
where evolution goes from the IR to the UV boundary $z_0$, i.e. the bulk action $S$ and Hamiltonian $H$ are taken at $z_0$.
Hamiltonian is
expressed through canonical momentum ${\mathcal \pi}$ conjugated to the gauge field ${\mathcal A}_0$ at the boundary
\begin{equation}
{\mathcal \pi}_{\alpha}=\frac{\partial L}{\partial(\partial_z{\mathcal A}_{0\alpha})}=\frac{\delta S}{\delta{\mathcal A}_{0\alpha}}
\label{}
\end{equation}
According to the AdS/CFT prescription, because ${\mathcal A}_0$ is a source of the current, we vary once and get
\begin{equation}
\langle j_{\alpha}\rangle = \frac{\delta S}{\delta{\mathcal A}_{0\alpha}}
\label{}
\end{equation}
From the action
\begin{equation}
S=\int d^2x dz \left( f^2((\partial_z A_{\mu})^2+(\partial_z V_{\mu})^2)-\frac{1}{2g^2}(F_{A\mu\nu}^2+F_{V\mu\nu}^2)
+4\kappa\varepsilon^{\nu\sigma}\partial_zV_{\nu}A_{\sigma}\right)
\label{action_1}
\end{equation}
we find the canonical momenta
\begin{equation}
\pi_{A\mu} =\frac{\partial L}{\partial(\partial_z A_{\mu})}=2f^2\partial_z A_{\mu},\quad
\pi_{V\mu} = \frac{\partial L}{\partial(\partial_z V_{\mu})}=2f^2\partial_z V_{\mu}+\phi_{V\mu}
\label{}
\end{equation}
where the shift in the canonical momenta due to the Chern-Simons term is 
\begin{equation}
\phi_{V\mu} = 4\kappa\varepsilon^{\mu\sigma}A_{\sigma}
\label{phi}
\end{equation}
and the corresponding 'velocities' are
\begin{equation}
\partial_zA_{\mu} = \frac{1}{2f^2} \pi_{A\mu},\,\,\partial_zV_{\mu} = \frac{1}{2f^2}{\tilde {\pi}_{V\mu}}
\label{}
\end{equation}
To simplify the notation we introduced shifted momentum ${\tilde {\pi}_{V\mu}}$,
\begin{equation}
\pi_{A\mu} =\langle j^5_{\mu}\rangle,\quad
{\tilde {\pi}_{V\mu}} = \pi_{V\mu}-\phi_{V\mu}=\langle j_{\mu}\rangle -\phi_{V\mu}
\label{tilde-pi}
\end{equation}
where $\phi$'s are given by eq.(\ref{phi}). This shows the mechanism how the bulk Chern-Simons term leads to the parity breaking in $1+1$D boundary
field theory . Namely the Chern-Simons term is responsible for the shift in canonical momenta which 
gives a nonzero vev of the current
\begin{equation}
\langle j_{\mu}\rangle = 4\kappa\varepsilon^{\mu\nu}{\mathcal A}_{\nu} \neq 0
\label{}
\end{equation}
Expressing velocities through momenta 
\begin{equation}
H = \int d^2x  (\pi_{A\mu}\partial_z A_{\mu}+ \pi_{V\mu}\partial_z V_{\mu} -L) 
\end{equation}
we obtain the Hamiltonian at the UV boundary, at $z=z_0$
\begin{eqnarray}
H &=& \int d^2 x \left(\frac{1}{2f^2}\pi_{A\mu}^2
+\frac{1}{2f^2}\pi_{V\mu}(\pi_{V\mu}-4\kappa\varepsilon^{\mu\sigma}A_{\sigma})\right.\nonumber\\
&-&\left.\left[f^2((\partial_zA_{\mu})^2+(\partial V_{\mu})^2)-\frac{1}{2g^2}(F_{A\mu\nu}^2+F_{V\mu\nu}^2)
+4\kappa\varepsilon^{\nu\sigma}\partial_z V_{\nu}A_{\sigma}\right]
\right)\nonumber\\
&=&\int d^2x \left(\frac{1}{4f^2}\pi_{A\mu}^2+\frac{1}{4f^2}(\pi_{V\mu}-\phi_{V\mu})^2
+\frac{1}{2g^2}(F_{A\mu\nu}^2+F_{V\mu\nu}^2)\right)
\label{hamiltonian}
\end{eqnarray}
Finally we arrive at the Hamilton-Jacobi equation (\ref{hamilton-jacobi}):
\begin{eqnarray}
\frac{\partial S}{\partial z_0} +\left.\int \frac{d^2q}{(2\pi)^2}\left(\frac{1}{4f^2}(\pi_{A\mu}^2 +\tilde{\pi}_{V\mu}^2)
+\frac{1}{2g^2}(F_{A\mu\nu}^2+F_{V\mu\nu}^2) \right)\right|^{z_0} =0
\label{hamilton-jacobi2}
\end{eqnarray}
where we traded time for holographic coordinate $z$ and the shifted momentum is given by eq.(\ref{tilde-pi}). 
Now, to obtain the corresponding RG equations for correlators
we vary the Hamilton-Jacobi eq.(\ref{hamilton-jacobi2}) with respect to boundary values of the gauge fields and introduce the following notation for the one-point
functions:
\begin{equation}
\pi_{A\alpha} =\frac{\delta S}{\delta A_{0\alpha}} = \langle j^5_{\alpha} \rangle,\quad
\pi_{V\alpha} =\frac{\delta S}{\delta V_{0\alpha}} = \langle j_{\alpha} \rangle
\end{equation}
and the two-point functions:
\begin{eqnarray}
 \frac{\delta^2S}{\delta A_{0\alpha}\delta A_{0\beta}} = \langle j^{5}_{\alpha}j^{5}_{\beta}\rangle
&=& \Pi_A \delta_{\alpha\beta},\quad
\frac{\delta^2S}{\delta V_{0\alpha}\delta V_{0\beta}} = \langle j_{\alpha}j_{\beta}\rangle
=\Pi_V \delta_{\alpha\beta},\\ 
 \frac{\delta^2S}{\delta V_{0\alpha}\delta A_{0\beta}} &=& \langle j_{\alpha}j^{5}_{\beta}\rangle
=\frac{1}{2\pi}\varepsilon_{\alpha\beta}w_T\equiv \varepsilon_{\alpha\beta}\tilde{w}_T
\label{currents}
\end{eqnarray}
which were calculated in section \ref{section:1}, we introduced a notation $w_T/(2\pi)=\tilde{w}_T$.
\\
\\
{\bf Hamilton-Jacobi equation for diagonal correlators}

First, we examine the RG equation for the diagonal correlators (\ref{vv}). We start off varying the Hamiltonian (\ref{hamiltonian}) twice with respect to the boundary value $A_0$, 
\begin{equation}
\frac{\delta^2}{\delta A_{0\alpha}\delta A_{0\beta}} \left.\left(\frac{1}{4f^2}(\pi_{A\mu}^2 +\tilde{\pi}_{V\mu}^2)
+\frac{1}{2g^2}(F_{A\mu\nu}^2+F_{V\mu\nu}^2) \right)\right|^{z_0}
\label{}
\end{equation}
we obtain
\begin{eqnarray}
&& \frac{1}{2f^2}\left( (\frac{\delta\pi_A}{\delta A_0})^2 + (\frac{\delta{\tilde {\pi}_V}}{\delta A_0})^2
+\pi_A\frac{\delta^2\pi_A}{\delta A_0^2}+{\tilde{\pi}_V}\frac{\delta^2\tilde{\pi}_V}{\delta A_0^2}\right)=\nonumber\\
&=& \frac{1}{2f^2}\left((\langle j^5 j^5\rangle)^2 + (\langle j j^5\rangle -\frac{\delta \phi_V}{\delta A_0})^2
+\langle j^5\rangle \langle j^5 j^5 j^5\rangle +(\langle j\rangle -\phi_V)\langle j j^5 j^5\rangle 
\right)=\nonumber\\
&=& \frac{1}{2f^2}\left(\Pi_A^2 +(\tilde{w}_T-4\kappa A|^{z_0})^2\right)\delta^{\alpha\beta}. 
\label{}
\end{eqnarray}
Our abelian action is quadratic
in fields, therefore we neglect three-point functions. Varying HJ eq.(\ref{hamilton-jacobi2}) and using eq.(\ref{condition}), 
we obtain the HJ equations for the diagonal correlators 
\begin{eqnarray}
&& \frac{\partial}{\partial z_0}\Pi_A +\frac{1}{2f^2}\left( \Pi_A^2 + (\tilde{w}_T-4\kappa)^2\right) = 0\\
&& \frac{\partial}{\partial z_0}\Pi_V +\frac{1}{2f^2}\left( \Pi_V^2 + (\tilde{w}_T-4\kappa)^2\right) = 0
\label{}
\end{eqnarray}
The HJ equation for the difference is given by
\begin{eqnarray}
\frac{\partial}{\partial z_0} (\Pi_A-\Pi_V) + \frac{1}{2f^2}(\Pi_A^2-\Pi_V^2) &=& 0, 
\label{hj-difference}
\end{eqnarray}
where all quantities are taken at the point $z_0$, i.e. $f(z_0)$, $\Pi_A(z_0)$ and $\Pi_V(z_0)$.
Using eq.(\ref{pi_lr}), we can rewrite eq.(\ref{hj-difference}) for the left-right correlator
\begin{eqnarray}
\frac{\partial}{\partial z_0} \Pi_{LR} &=& -\frac{1}{2f^2}(\Pi_A+\Pi_V)\Pi_{LR}.
\label{hj-pi_lr}
\end{eqnarray}
The RG equation for the left-right correlator is diagonal, i.e. its running is expressed again through the left-right correlator. The momentum dependent coefficient is given by the sum of the correlators $\Pi_A+\Pi_V$. 
\\
\\
{\bf Hamilton-Jacobi equation for mixed correlator}

Varying the Hamiltonian part in the HJ eq.(\ref{hamilton-jacobi2}) twice with respect to the boundary values $V_0$ and $A_0$, 
we obtain
\begin{equation}
\frac{\delta^2}{\delta V_{0\alpha}\delta A_{0\beta}}\left.\left(\frac{1}{4f^2}(\pi_{A\mu}^2 +\tilde{\pi}_{V\mu}^2)
+\frac{1}{2g^2}(F_{A\mu\nu}^2+F_{V\mu\nu}^2) \right)\right|^{z_0}
\end{equation}
we obtain
\begin{eqnarray}
&& \frac{1}{2f^2}\left(\frac{\delta\pi_A}{\delta V_0}\frac{\delta\pi_A}{\delta A_0}  
+\frac{1}{2f^2}\frac{\delta\tilde{\pi}_V}{\delta V_0}\frac{\delta\tilde{\pi}_V}{\delta A_0}
+\pi_A\frac{\delta^2\pi_A}{\delta V_0\delta A_0}
+\tilde{\pi}_V\frac{\delta^2\tilde{\pi}_V}{\delta V_0\delta A_0}
\right)=\nonumber\\
&=& \frac{1}{2f^2}\left(\langle j j^5\rangle \langle j^5 j^5\rangle
+(\langle j j^5\rangle -\frac{\delta \phi_V}{\delta A_0})\langle j j\rangle
+\langle j^5\rangle \langle j j^5 j^5\rangle + (\langle j\rangle -\phi_V)\langle j^5 j j\rangle
\right)=\nonumber\\
&=& \frac{1}{2f^2}\left(\varepsilon^{\alpha\gamma}\tilde{w}_T\Pi_A\delta^{\gamma\beta}
+(\left.\varepsilon^{\gamma\beta}\tilde{w}_T-\varepsilon^{\gamma\beta}4\kappa A\right|^{z_0})\Pi_V\delta^{\gamma\alpha}\right)=\nonumber\\
&=& \frac{1}{2f^2}\varepsilon^{\alpha\beta}\left(\tilde{w}_T\Pi_A+(\tilde{w}_T-4\kappa)\Pi_V\right).
\label{}
\end{eqnarray}
Varying HJ eq.(\ref{hamilton-jacobi2}), 
we obtain the HJ equations for the mixed correlator 
\begin{eqnarray}
\frac{\pr}{\pr z_0}\tilde{w}_T+\frac{1}{2f^2}\left(\tilde{w}_T(\Pi_A+\Pi_V)+ 2\kappa(\Pi_A-\Pi_V)-2\kappa(\Pi_A+\Pi_V)\right)=0,
\label{hj-mix}
\end{eqnarray}
where $f(z_0)^2$, $\tilde{w}_T(z_0)$, $\Pi_A(z_0)$ and $\Pi_V(z_0)$ are taken at $z_0$. As seen from eq.(\ref{hj-mix}),
due to the Chern-Simons term, $\kappa\neq 0$, the RG equation for the mixed correlator $\tilde{w}_T$ is not diagonal
\begin{eqnarray}
\frac{\pr}{\pr z_0}\tilde{w}_T=-\frac{1}{2f^2}(\Pi_A+\Pi_V)\tilde{w}_T + \frac{2\kappa}{2f^2}(\Pi_A+\Pi_V)-\frac{2\kappa}{2f^2}(\Pi_A-\Pi_V)
\label{}
\end{eqnarray}
It is remarkable that the diagonal RG flow for $w_T$ has the same rate $(1/2f^2)(\Pi_A+\Pi_V)$ as 
the left-right correlator $\Pi_{LR}$.
\\
\\
{\bf Hamilton-Jacobi equation for Son-Yamamoto relation}

We write the HJ equation for the Son-Yamamoto relation eq.(\ref{son-yamamoto-relation})
\begin{equation}
(SY)= \tilde{w}_T-2\kappa+2\kappa\int^{z_0} \frac{dz}{2f^2}(\Pi_A-\Pi_V),
\label{sy3}
\end{equation}
where $\tilde{w}_T=w_T/(2\pi)$.
To this end we differentiate eq.(\ref{sy3}) with respect to $z_0$ and use the HJ equations for the diagonal
and mixed two-point functions, eqs.(\ref{hj-difference},\ref{hj-mix}),
\begin{eqnarray}
&&\frac{\partial }{\partial z_0}\tilde{w}_T +2\kappa\int^{z_0}\frac{dz}{2f^2}\frac{\partial}{\partial z_0}(\Pi_A-\Pi_V) +\frac{2\kappa}{2f^2}(\Pi_A-\Pi_V)+\\
&+&\frac{1}{2f^2}\left(\tilde{w}_T(\Pi_A+\Pi_V)+2\kappa(\Pi_A-\Pi_V)-2\kappa(\Pi_A+\Pi_V)\right)
+2\kappa\int^{z_0}\frac{dz}{2f^2}\frac{1}{2f^2}(\Pi_A^2-\Pi_V^2)-2\kappa\frac{1}{2f^2}(\Pi_A-\Pi_V)=0,\nonumber\\
\label{}
\end{eqnarray}
where the first and the fourth terms constitute eq.(\ref{hj-mix}), and the second and the fifth terms constitute eq.(\ref{hj-difference}).
Note that the integral with the metric term $1/f^2$ in eq.(\ref{sy3}) is differentiated.
Combining the terms we have
\begin{eqnarray}
\frac{\partial}{\partial z_0}\left(\tilde{w}_T-2\kappa+2\kappa\int\frac{dz}{2f^2}(\Pi_A-\Pi_V)\right)
+\frac{1}{2f^2}(\Pi_A+\Pi_V)\left(\tilde{w}_T-2\kappa+2\kappa\int\frac{dz}{2f^2}(\Pi_A-\Pi_V)\right)=0,
\label{}
\end{eqnarray}
that can be written in a short form
\begin{eqnarray}
\frac{\partial}{\partial z_0}(SY)=-\frac{1}{2f^2}(\Pi_A+\Pi_V)(SY)
\label{son-yamamoto-rg}
\end{eqnarray}
where $(SY)$ denotes the Son-Yamamoto relation eq.(\ref{sy3}). The RG flow for the Son-Yamamoto relation is diagonal. 
It is remarkable, that the Son-Yamamoto relation $(SY)$ and the left-right correlator $\Pi_{LR}$ eq.(\ref{hj-pi_lr}) both flow with the same coefficient which is given by the sum $\sim (\Pi_A+\Pi_V)$.

In section \ref{section:4} (the regime of small momenta) we showed that the Son-Yamamoto relation eq.(\ref{sy3}) is satisfied 
at the point $z_0\rightarrow 0$. This means that since the RG eq.(\ref{son-yamamoto-rg}) is diagonal,  
the Son-Yamamoto relation holds for any energy scale $z_0$.

\section{Similarity of QCD and two-dimensional system. Dimensional reduction}\label{section:3}

In this section we draw parallels between the $4$-dimensional QCD \cite{son-yamamoto},\cite{dubinkin-gorsky-milehin2015} 
and our $2$-dimensional system. We write formulas for the $2$D system in the context of ${\rm QCD}_2$. 
We summarize the RG equations 
\begin{eqnarray}
\frac{\partial}{\partial z_0} \Pi_{LR} &=& -\frac{1}{2f^2}(\Pi_A+\Pi_V)\Pi_{LR}\\
\frac{\partial}{\partial z_0}(SY) &=& -\frac{1}{2f^2}(\Pi_A+\Pi_V)(SY)
\label{}
\end{eqnarray}
which are identical in both QCD and linear cases. It is remarkable that the two equations have the same rate of change 
$\frac{1}{f^2}\left(\Pi_A+\Pi_V \right)$. Further comparing ${\rm QCD}_4$ and ${\rm QCD}_2$, the polariztion operators $\Pi_A$ and $\Pi_V$ are the same, however the Son-Yamamoto relations slightly differ. Explicitly they are given by
\begin{eqnarray}
2{\rm D}:\,(SY)&=& w_T -N_c+ N_c\int_0^{z_0} \frac{dz}{2f^2}(\Pi_A-\Pi_V)=0,\\
\langle j_{\mu}j_{\nu}^5\rangle^{\perp} &=& \frac{1}{2\pi}\left(N_c-N_c\int_0^{z_0}\frac{dz}{2f^2}(\Pi_A-\Pi_V)\varepsilon_{\mu\nu}\right),
\label{sy1}
\end{eqnarray}
where $w_T\sim \langle j_A(q) j_V(-q)\rangle^{\perp}$,
and \cite{son-yamamoto}
\begin{eqnarray}
4{\rm D}:\,(SY)&=& w_T-\frac{N_c}{Q^2}+\frac{N_c}{f_{\pi}^2}(\Pi_A-\Pi_V)=0,\\
\langle j_{\mu}j_{\nu}^5\rangle^{\perp} &=& \frac{Q^2}{4\pi^2}\left(\frac{N_c}{Q^2}-\frac{N_c}{f_{\pi}^2}(\Pi_A-\Pi_V)\tilde{F}_{\mu\nu}\right),
\label{sy2}
\end{eqnarray}
where the dual field strength $\tilde{F}_{\mu\nu}=1/2\varepsilon_{\mu\nu\alpha\beta}F^{\alpha\beta}$, and
$w_TF(k)\sim \langle j_A(q)j_V(-q-k)j_V(k)\rangle^{\perp}$ with $k=0$, $F$ is the field strength of the vector gauge field, $Q^2=-q^2$. 
The following identification is done
\begin{equation}
\frac{1}{f_{\pi}^2}=\int_0^{z_0} \frac{dz}{2f^2}.
\label{definition}
\end{equation}
Note that the dimension of the pion decay constant is $[f_{\pi}]=1$ in $2$D and $[f_{\pi}]=E$ in $4$D.
     
Next we consider the dimensional reduction $d\rightarrow d-2$ that occurs at strong magnetic field $B\rightarrow\infty$, 
in order to see a connection between $4$D QCD and $2$D systems. 
Dirac action is written as
\begin{equation}
S_{F}=i\int d^4x \bar{\psi}\left(\Gamma_{\mu}D^{\mu}-m\right)\psi
\label{}
\end{equation}
where $\Gamma$ are the four component gamma matrices, and the covariant derivative contains the gauge field.  
We choose the gauge
\begin{equation}
A_y=-yB
\label{}
\end{equation}
with $B\parallel z$ and $B$ is positive, and consider a z-slice for the time being. Then we decompose
the Dirac spinor into two two-component Weyl spinors  
\begin{eqnarray}
\psi= {\rm e}^{-i\omega t +ikx}\left(
\begin{array}{c}
 \xi_1(y)\\
 \xi_2(y)
\end{array}
\right)
\label{}
\end{eqnarray}
with $k_x\equiv k$. For a new variable
\begin{equation}
\eta=\sqrt{B}\left(y+\frac{k}{B}\right)
\label{}
\end{equation}
the Dirac equation for $\xi_i$ is reduced to a harmonic oscillator, where a solution is defined in terms of the Hermite polynomials $H_n$
\begin{eqnarray}
&& \xi_1= c(\omega,k) I_n(\eta), \quad \xi_2= \pm c(\omega,k) I_{n-1}(\eta) \nonumber\\
&& I_n(\eta)=\frac{1}{\sqrt{2^nn!\sqrt{\pi}}}{\rm e}^{-\eta^2/2}H_n(\eta)
\label{}
\end{eqnarray}
and the energy is quantized 
\begin{equation}
\omega=\pm \sqrt{2Bn}
\label{}
\end{equation}
$n=0,1,2,\cdots$ are Landau levels; where $\pm$ distingiushes the two solutions. Motion in perpendicular direction to the magnetic field $(x,y)$
is described by Larmor orbits. 
In the limit $B\rightarrow\infty$, only the lowest Landau level (LLL)
is important. Indeed the LLL
has a vanishing energy, because the zero point energy $\frac{1}{2}B$ is exactly compensated by the Zeeman splitting due
to spin coupling $-\frac{1}{2}B$. Therefore the zero modes for each of the two-component spinors
is given by 
\begin{eqnarray}
\xi_i={\rm e}^{-\eta^2/2}\left(
\begin{array}{c}
0\\
\zeta_i
\end{array}
\right)
\label{}
\end{eqnarray}
The fact that only one spin component is populated means that the LLL is spin polarized. Reinstating the $z$ dependence back, the zero modes
become functions $\zeta_i(t,z)$. There is one zero mode for each state of the LLL, for each $k$. These zero modes are described by $1+1$-dimensional
effective action for a two-component Weyl spinor $\zeta=(\zeta_1,\zeta_2)$  
\begin{equation}
S_{eff}=i\int d^2x \bar{\zeta}\gamma_{\mu}D_{\mu}\zeta
\label{}
\end{equation}
where $\gamma$ are given by the Pauli marices, and the covariant derivative does not contain the gauge field now. 
In strong magnetic fields $B\rightarrow \infty$ where only the LLL is important, the dynamics is reduced from $4$D to $2$D.
Since the LLL is spin polarized, the density of states for the LLL is $\frac{B}{2\pi}$. This means that in the limit $B\rightarrow\infty$ in order to get
one- and two-point functions of the currents, we calculate correlators for the two-dimensional fermions, and then sum over the fermi zero modes using the density of states in the LLL   
\begin{equation}
\langle \bar{\psi}\Gamma\psi\rangle =\frac{B}{2\pi}\langle \bar{\zeta}\gamma\zeta\rangle
\label{}
\end{equation}
and schematically 
\begin{equation}
\langle J(x)J(0)\rangle_{4d} =\frac{B}{2\pi}\langle j(x)j(0)\rangle_{2d}
\label{}
\end{equation}
where $\bar{\psi}\Gamma\psi = J$ and $\bar{\zeta}\gamma\zeta = j$ are fermion currents in $4$D and $2$D, respectively.
Similar calculations can be done in a holographically dual theory with dual fermions and currents, where the reduction
in the bulk theories $5$D to $3$D occurs \cite{albash-johnson,bolognesi-tong}. This means that at large $B$ the dimensional reduction  
$4$D to $2$D for the current correlators holds also nonperturbatively.

\section{Conclusions}

We derived the analog of Son-Yamamoto relation for (1+1)-dimensional systems. Two dimensional systems are presently realized by organic quasi-$1$D metals, organic nanotubes, edge states of quantum Hall liquids, $1$D semiconducting structures, 
and edge states of topological insulators \cite{one-dim-system}. In these systems,
it is believed that electron-electron interaction invalidates Landau Fermi liquid picture. Instead a different state described approximately by 
Tomanaga-Luttinger theory \cite{tomanaga-luttinger} is generated. Since electronic correlations in this state are stronger than in Fermi liquid,
it is interesting to calculate two-point correlations that represent conductivities or related transport coefficients and obtain relations between them. 
For example it is interesting to translate our transport coefficients in terms of the Coulomb/spin drag trans-resistivity between two quantum wires \cite{coulomb-drag}; or examine transport properties in chiral edge states in quantum Hall state and helical edge states 
in topological insulators/topological superconductors \cite{top-ins/sup}.

We summarize the two representations of the Son-Yamamoto relation for $(1+1)$-dimensional systems.
The one that relates $w_T$ -- the mixed $\langle VA\rangle$ current correlator and the diagonal $\langle VV\rangle$ 
and $\langle AA\rangle$ current correlators 
\begin{equation}
(SY)= w_T -4\pi\kappa+4\pi\kappa\int_0^{z_0} \frac{dz}{2f^2}(\Pi_A-\Pi_V)=0,
\label{}
\end{equation}
and the other written for the left-right correlator $\langle LR\rangle$
\begin{equation}
\langle j_{\mu}^L j_{\nu}^R\rangle^{\perp} = \varepsilon_{\mu\nu}\left(4\kappa-4\kappa\int_0^{z_0}\frac{dz}{2f^2(z)}\Pi_{LR}\right),
\label{}
\end{equation}
where $j_{\mu}^L$ and $j_{\mu}^R$ are the left- and right-handed currents. In the context of the ${\rm QCD}_2$, $4\pi\kappa=N_c$.
The key point in deriving the Son-Yamamoto relation was the independence on the radial coordinate of the Wronskian for vector and axial gauge fields.
It gives the range of validity for the Son-Yamamoto: small Chern-Simons $\kappa$ or large virtuality $Q^2$. It would be instructive to get
qualitative estimates for this parameter range.

The two-dimensional Son-Yamamoto matching condition at large virtualities 
provides an estimate for the decay constant
\begin{equation}
f_{\pi}^2\sim N_c,
\label{estimate2}
\end{equation}  
which is found in the limit of the weak coupling $N_c\rightarrow \infty$ and 't Hooft condition $g^2N_c={\rm const}$
by Zhitnitsky \cite{zhitnitsky}. 
Since this estimate is done at large $Q^2$ where the application of the low-energy 
effective action is questionable this result deserves for the independent derivation by other means.
We also showed that the pion decay constant $f_{\pi}^2\sim N_c$ is consistent with the Gell-Mann-Oakes-Renner relation
and the chiral condensate $\langle\bar{\psi}\psi\rangle\sim N_c$.  
In ${\rm QCD}_4$, the analog of the estimate for $f_{\pi}$ eq.(\ref{estimate2}) is the holographic result for magnetic susceptibility of Vainstein \cite{Vainshtein2002}
$\chi\sim 1/f_{\pi}^2$ \cite{son-yamamoto}.

Finally we found, that the RG flow equations for the Son-Yamamoto relations in $(1+1)$ and $(3+1)$-dimensional systems are the same and they
are diagonal. Moreover, the rate of the RG flow for SY relation and the left-right correlator is the same
\begin{eqnarray}
\frac{\partial}{\partial z_0} \Pi_{LR} &=& -\frac{1}{2f^2}(\Pi_A+\Pi_V)\Pi_{LR}\\
\frac{\partial}{\partial z_0}(SY) &=& -\frac{1}{2f^2}(\Pi_A+\Pi_V)(SY)
\label{}
\end{eqnarray}
where $z_0$ is the UV boundary value of the radial bulk coordinate - the end point of the evolution. We believe that the diagonal form and this rate
holds only for the abelian case.
We showed that the similarity between $(3+1)$D and $(1+1)$D systems can be attributed to the dimensional reduction
$D\rightarrow D-2$ in strong magnetic field. However, it does not explain why the RG flows are diagonal
and have the certain rate.

\section*{Acknowledgements}

The authors thank Pavel Buividovich, Maxim Chernodub, Alexey Milekhin, Valentin Zakharov for discussions. This work is supported, in part, by grants 
RBBR-15-02-02092 (A.G,O.R) and Russian Science Foundation grant for IITP
14-050-00150 (A.G). A.G. thanks SAITP and San Paulo University where the part of the work has been done
for the hospitality and support.

\clearpage
\appendix

\section{Checking the Son-Yamamoto relation in a model. 1+1 systems in an AdS model with the chiral condensate}\label{appendix:b}

We consider Son-Yamamoto relation for 1+1 systems in a gravity dual model which incorporates the chiral condensate \cite{erlich-katz-son-stephanov}.
Contrary to \cite{erlich-katz-son-stephanov}, we do not impose the hard-wall cutoff in the IR that insured confinement in 3D QCD. 
The metric is a slice of the AdS space 
\begin{equation}
ds^2=\frac{1}{z^2}\left(-dt^2+dz^2+dx^2\right),
\end{equation}
where $0 \leq z \textless \infty$, the AdS UV boundary is at $z=0$, and we rescale the curvature radius of the 
space to unity. 
The action in the bulk 
\begin{equation}
S = S_{YM}+S_{CS} 
\end{equation}  
includes the scalar field
\begin{eqnarray}
S_{YM} &=& \int d^3 x \sqrt{g} \left(|D\Psi|^2+M^2|\Psi|^2-\frac{1}{4g_3^2}(F_L^2+F_R^2)\right), \label{gauge_action}\\
S_{CS} &=& \kappa \int d^3 x \left(w_3(A_L)-w_3(A_R)\right),\label{Chern-Simons_action} 
\end{eqnarray} 
where $D\Phi=\partial \Phi -iA_L\Phi +iA_R\Phi$, $F_{\mu\nu}=\partial_{\mu}A_{\nu}-\partial_{\nu}A_{\mu}$, 
$w(A)=A\ast F +\frac{2}{3}A^3$, and $M^2$ is specified further. From the AdS dictionary, a bulk field $\Theta$ dual to operator $O$ behaves at the asymptotic
UV boundary ${\rm AdS}_{d+1}$ as
\begin{equation}
\Theta(z)={\mathcal A}z^{\Delta_{-}}(1+\cdots)+{\mathcal B}z^{\Delta_{+}}(1+\cdots), \;\;z\to 0, 
\end{equation}   
where the source to $O$ (leading term) is ${\mathcal A}$, the expectation of $O$ (subleading term) is ${\mathcal B}=\langle O\rangle$,
and the characteristic exponents $\Delta_{\pm}$ for scalar and vector fields are solutions to equations
\begin{eqnarray}
&& {\rm scalar}:\;\;\Delta(\Delta-d)=M^2,\\
&& {\rm vector}:\;\;\Delta(\Delta-d+2)=M^2,
\label{characteristic_exp}
\end{eqnarray} 
where the AdS curvature radius is one. We take the scalar mass equal to the Breitenlohner-Freedman (BF) bound $M^2=-1$ to insure the positive energy, 
and the mass of vector field is $M^2=0$. From eq.{\ref{characteristic_exp}}, for the ${\rm AdS}_3$, $d=2$ 
the characteristic exponents are $\Delta_{\pm}=1$ for scalar
and $\Delta=0$ for vector fields. This implies the following behavior in the UV
\begin{eqnarray}
&& {\rm scalar}:\;\;\Psi=mz\ln z +\langle q\bar{q}\rangle z, \label{scalar}\\
&& {\rm vector}:\;\; V= A\ln z +\langle J\rangle.\label{vector}
\end{eqnarray} 
In the context of QCD, the source is the quark mass $m$ and the response is
the chiral condensate $\langle q\bar{q}\rangle$, and the EM field $A$
sources the $U(1)$ conserving current with expectation value $\langle J\rangle$.  
In order to check that the scalar field behaves as in eq.(\ref{scalar}), we solve the equation of motion for $\Psi$ 
without the gauge field
\begin{equation}
\partial_z\left(\frac{1}{z}\partial_z\Psi\right)-\frac{1}{z^3}M^2\Psi=0. 
\end{equation}  
Indeed, the solution is 
\begin{equation}
\Psi =\frac{1}{2}m z\ln z +\frac{1}{2}\sigma z,
\label{source} 
\end{equation}
with $M^2=-1$, and $\sigma$ is the chiral condensate. As in \cite{Krikun2008}, we parametrize the scalar field as
\begin{equation}
\Psi=\Psi_0{\rm e}^{i2\pi},\;\;\Psi_0=\frac{1}{2}v(z),\;\;v(z)=mz\ln z+\sigma z.\label{parametrize_v(z)} 
\end{equation}
Introducing the vector and axial-vector fields, $V=(A_L+A_R)/2$ and $A=(A_L-A_R)/2$,
the covariant derivative for the scalar becomes $D\Psi =2 i \Psi_0(\partial \pi - A)$. We work in the radial gauge,
$V_z=A_z=0$. We decompose the gauge fields as
\begin{eqnarray}
V_{\mu} &=& V^{\perp}_{\mu},\;\; \partial_{\mu}V_{\mu}=0,\\
A_{\mu} &=& A^{\perp}_{\mu}+A^{\parallel}_{\mu},\;\; \partial_{\mu}A^{\perp}_{\mu}=0. 
\end{eqnarray}   
The action eq.(\ref{gauge_action}) $S_{YM}=S_{V}+S_{A}$ is
\begin{eqnarray}
S_{V} &=& \int d^3x \left(-\frac{1}{4g_3^2}\right)2z F_V^2,\\
S_{A} &=& \int d^3x \left[\left(-\frac{1}{4g_3^2}\right)2z F_A^2+\frac{v^2(z)}{z}(\partial\pi-A)^2\right],
\end{eqnarray}
that can be written as 
\begin{eqnarray}
S_{V} &=& \int d^3x \left(-\frac{1}{4g_3^2}\right)\left[2z F^{\perp\,2}_{V\mu\nu} + 4z F^{\perp\,2}_{Vz\mu}\right],\label{gauge_action_V}\\
S_{A} &=& \int d^3x \left(-\frac{1}{4g_3^2}\right)\left[2z F^{\perp\,2}_{A\mu\nu} + 4z F^{\perp\,2}_{Az\mu}
+ 4z F^{\parallel\,2}_{Az\mu}\right]\nonumber\\
&+&\int d^3x \frac{v^2(z)}{z}\left[(\partial_z\pi-A_z)^2+(\partial_{\mu}\pi-A_{\mu}^{\parallel})^2+A_{\mu}^{\perp\,2}
\right],\label{gauge_action_A} 
\end{eqnarray}
where $v(z)$ is given by eq.(\ref{parametrize_v(z)}).
Comparing the gauge action in eq.(\ref{yang-mills2}) and eqs.(\ref{gauge_action_V},\ref{gauge_action_A}), the following identification of the metric factors
in eq.(\ref{yang-mills2}) can be made 
\begin{equation}
f^2(z)=\frac{1}{4g_3^2}\frac{2\sqrt{g}}{g_{zz}g_{\mu\mu}}=\frac{z}{2g_3^2},\;\;\frac{1}{g^2}=\frac{1}{4g_3^2}\frac{2\sqrt{g}}{g^2_{\mu\mu}}=\frac{z}{2g_3^2},
\label{metric-factor}
\end{equation}
where the determinant is $\sqrt{g}=1/z^3$ for the ${\rm AdS}_3$, (not to confuse factor $1/g^2$ in eq.(\ref{yang-mills2}) with the metric determinant $g$).
Let the gauge fields are $V^{\mu}(q,z)=V(q,z)V_0^{\mu}$ and $A^{\mu}(q,z)=A(q,z)A_0^{\mu}$ with $V_0,A_0$ being sources of the vector and axial-vector currents,
and $q$ is the Fourier transform momentum in the boundary space component $x$. 
From eqs.(\ref{gauge_action_V},\ref{gauge_action_A}), the linearized equations of motion (EOM) for the perpendicular components of the vector and axial-vector fields,
$V(q,z)$ and $A(q,z)$, read
\begin{eqnarray}
&& \partial_z(z\partial_z V) - zQ^2V = 0,\\
&& \partial_z(z\partial_z A) - zQ^2A -\frac{g_3^2v^2}{z}A = 0,
\end{eqnarray}  
where $Q^2=-q^2$, and we ommit the perpendicular sign. The boundary conditions (BC) in the UV and IR are
\begin{eqnarray}
&{\rm UV:}&\;\;z\partial_z V|_{z=0}=1,\;\; z\partial_z A|_{z=0}=1,\label{boundary_condition_UV}\\
&{\rm IR:}&\;\;\partial_z V|_{z_m\rightarrow\infty}=0,\;\; \partial_z A|_{z_m\rightarrow\infty}=0,\label{boundary_condition_IR}
\end{eqnarray}  
where we introduced the hard-wall cutoff $z_m$ which we let to infinity. The UV BC says that the source for the components $V(q,z)$ and $A(q,z)$ is unity.
Indeed the source of the vector field is given by $zV'_{z=0}$ when asymptotic behavior is as in eq.{\ref{vector}}. 
First, we solve the EOM for the vector field
\begin{equation}
z^2V^{\prime\prime}+zV^{\prime}-Q^2z^2V=0.
\end{equation} 
The solution 
\begin{equation}
V=c_1I_0(Qz)+c_2K_0(QZ)
\end{equation} 
is expressed through the modified Bessel functions $I_n,K_n$ with $n=0$. Imposing BC's
\begin{eqnarray}
&{\rm IR:}&\;\;c_1I_1(Qz_m)-c_2K_1(Qz_m)=0,\; z_m\rightarrow\infty\\
&{\rm UV:}&\;\;zQ\left(c_1I_1(0)-c_2K_1(0)\right)=1,
\end{eqnarray}  
we obtain
\begin{equation}
V_{\perp}(Q,z)=-K_0(Qz)-\frac{K_1(Qz_m)}{I_1(Qz_m)}I_0(Qz)\xrightarrow{z_m\rightarrow\infty} -K_0(Qz).
\label{solution_V}
\end{equation} 
Using asymptotic expansion at $z=0$ for the modified Bessel functions 
\begin{equation}
I_0(z)\approx 1+z^2/4 +\cdots,\;\; K_0\approx (-\gamma+(1-\gamma)z^2/4+\cdots)-\ln (z/2)(1+z^2/4+\cdots),
\end{equation}
we find that $V$ behaves in the UV as in eq.(\ref{vector}):
\begin{equation}
V_{\perp}(Q,z)\rightarrow \ln(Qz) + {\rm const},\; z\rightarrow 0
\label{solution-asymptotic} 
\end{equation}   
with the source being unity.

Next, we solve the EOM for the perpendicular component of the axial-vector field perturbatively for large $Q^2\rightarrow\infty$
\begin{equation}
A=A_0+A_1+\cdots,
\end{equation} 
with $A_0(Q,z)=V_{\perp}(Q,z)$ eq.(\ref{solution_V}). The first correction satisfies the equation
\begin{equation}
x^2\partial_x^2A_1+x\partial_xA_1-xA_1=\lambda x^2A_0,
\label{equation_A}
\end{equation}
where we defined $x=Qz$, $\lambda =g_3^2\sigma^2/Q^2$ and $\lambda\rightarrow 0$ as $Q\rightarrow\infty$. The solution of this equation can be found by using
the Green's function
\begin{equation}
A_1=\int dx^{\prime}G(x,x^{\prime})\lambda x^{\prime\,2}A_0(x^{\prime}),
\end{equation}
where the Green's function is obtained from solving the homogeneous part of equation (\ref{equation_A})
\begin{equation}
f_1=-K_0(x),\;\; f_2(x)=-I_0(x),
\end{equation}
and 
\begin{equation}
G(x,x^{\prime})=-\frac{1}{W[f_1,f_2](x^{\prime})}\left(f_1(x)f_2(x^{\prime}\Theta(x-x^{\prime})+f_2(x)f_1(x^{\prime})\Theta(x^{\prime}-x)\right),
\end{equation}
with the Wronskian 
\begin{equation}
W[f_1,f_2](x)=f_1f_2^{\prime}-f_1^{\prime}f_2=1/x 
\end{equation}
which is simple to estimate for $x\rightarrow 0$. We find the small $z$ behavior of $A_1$
\begin{equation}
A_1(Q,z)=\frac{1}{3}\frac{g_3^2\sigma^2}{Q^2},
\end{equation}
where we used the following intergral
\begin{equation}
\int_{0}^{\infty}dxx^{3}K^2_0(x)=\frac{1}{3}.
\end{equation}
Summarizing, solutions near the boundary for vector and axial-vector fields are
\begin{eqnarray}
V_{\perp}(Q,z) &=& -K_0(Qz),\\
A_{\perp}(Q,z) &=& -K_0(Qz) + \frac{1}{3}\frac{g_3^2\sigma^2}{Q^2}.
\end{eqnarray}  
As pointed out in \cite{son-yamamoto}, these solutions near the boundary are sufficient to evaluate the 2-point correlation functions below
which are determined by the boundary values at $z=\epsilon$ or by the intergrals dominated by small $z$ regions.

The derivation of correlation functions is similar to the one in section \ref{section:1}. Therefore we put the resulting expressions here.
The transverse parts of the diagonal vector and axial current correlation functions are
\begin{eqnarray}
\Pi_V(Q^2) &=& -\frac{1}{g_3^2}V_{\perp}(Q,z)|_{z=\epsilon},\\
\Pi_A(Q^2) &=& -\frac{1}{g_3^2}A_{\perp}(Q,z)|_{z=\epsilon},
\end{eqnarray} 
where we used the boundary condition $zV^{\prime}|_{z=\epsilon}=zA^{\prime}|_{z=\epsilon}=1$. 
We introduce a cutoff $\Lambda$ as $\epsilon=1/\Lambda$.
The transverse part of the mixed
vector-axial current correlator is 
\begin{equation}
w_T = 2\kappa\int_{0}^{z_m\rightarrow\infty} dz \left(A^{\prime}(Q,z)V(Q,z)-V^{\prime}(Q,z)A(Q,z)\right).
\end{equation}
Here we don't add the boundary term eq.(\ref{chern-simons200}) because the gauge fields diverge on the boundary.
Expanding $V_{\perp}$ and $A_{\perp}$ near the boundary, 
\begin{eqnarray}
V_{\perp} &\approx& \frac{1}{2}\ln (Q^2z^2)+{\rm const} +O(z^2),\\
A_{\perp} &\approx& \frac{1}{2}\ln (Q^2z^2)+ \frac{1}{3}\frac{g_3^2\sigma^2}{Q^2}+{\rm const} +O(z^2),
\end{eqnarray} 
with ${\rm const}=\gamma-\ln 2$,
we obtain for the diagonal and mixed current correlators
\begin{eqnarray}
\Pi_V(Q^2) &=& -\frac{1}{2g_3^2}\ln Q^2,\label{piV-solution}\\
\Pi_A(Q^2) &=& -\frac{1}{2g_3^2}\ln Q^2 - \frac{1}{3}\frac{\sigma^2}{Q^2},\label{piA-solution}\\
w_T(Q^2) &=& -2\kappa \frac{1}{3} \frac{g_3^2\sigma^2}{Q^2} \ln\Lambda, \label{wT-solution}
\end{eqnarray}
where we used for evaluating $w_T$:
\begin{equation}
\int_0^{\infty}dx K_0^{\prime}(x)=K_0(x)|_{x=\epsilon}=-\ln\Lambda,
\end{equation}
with asymptotic value $K_0(x)=-\ln x$ at small $x$, and $\epsilon=1/\Lambda$.
Combining the above results, we obtain
\begin{equation}
w_T(Q^2)=2\kappa g_3^2\left(\Pi_A(Q^2)-\Pi_V(Q^2)\right)\ln\Lambda.
\end{equation}
This expression should be compared with the Son-Yamamoto relation eq.(\ref{son-yamamoto-relation}). Using eq.(\ref{metric-factor}) for the metric factor,
the integral in Son-Yamamoto relation becomes
\begin{equation}
\int_0^{\infty}\frac{dz}{2f^2(z)}=\int_{z=\epsilon} dz \frac{g_3^2}{z}=g_3^2\ln\Lambda,
\end{equation}
which has the same divergent leading log behavior as $w_T$. It proves that Son-Yamamoto relation eq.(\ref{son-yamamoto-relation}) is satisfied.


Now we calculate the left-right correlator $\Pi_{LR}$ eq.(\ref{pi_lr}), to which
the terms proportional to the chiral condensate and the fermion mass
contribute. We consider the chiral limit with zero fermion mass $m=0$.  
Using eqs.(\ref{piV-solution},\ref{piA-solution}), we find that the left-right correlator is     
\begin{equation}
\Pi_{LR}= -\frac{1}{3}\frac{\sigma^2}{Q^2}
\end{equation}
up to the quadratic order in the chiral condensate. 

Finally let us evaluate the chiral condensate that  
fixes the relation between $\sigma$ and $\langle\bar{\psi}\psi\rangle$. 
In the field theory one can evaluate the condensate as the variation of the vacuum energy
with respect to the quark mass. In the dual theory, because $m$ is the source for $\sigma$ eq.(\ref{source}), 
we variate action on the classical solution \cite{Krikun2008}
\begin{equation}
\langle\bar{\psi}\psi\rangle =\left.\frac{\delta S_{A}(\Psi_0)}{\delta m}\right|_{m=0},
\label{}
\end{equation}
where the action is given in eq.(\ref{gauge_action_A}). The variation of the action is
\begin{equation}
\delta S_A = \int d^2x \left.\sqrt{g}4\partial_z\Psi\delta\Psi\right|_{z=0}=\int d^2x \frac{2}{z} (m\ln z +m +\sigma)z\delta m |_{z=0},
\label{}
\end{equation}
where we used eq.(\ref{parametrize_v(z)}).
Therefore the chiral condesate is
\begin{equation}
\langle\bar{\psi}\psi\rangle =2\sigma.
\label{}
\end{equation}

\section{Numerical solution for the "cosh" and Sakai-Sugimoto models}\label{appendix:b}

In this appendix, we solve equations of motion numerically and show that Wronskian is independent on the radial coordinate
in the regime of large momentum $Q^2$ and at large radial coordinate $z$. As we showed in Appendix A, the $3$D bulk solutions are divergent.
Using the analog of $(4-\epsilon)$ dimensional regularization, we will work in the bulk $(3+\epsilon)$ dimensions and take the limit 
of small $\epsilon$ at the end.
Let us start with the equations of motion:
\begin{eqnarray}
\partial_z(f^2(z)\partial_zA_{\mu})-\frac{Q^2}{g^2}A_{\mu} + 2 \kappa\varepsilon_{\mu\nu}\partial_zV_{\nu} &=& 0,\nonumber\\
\partial_z(f^2(z)\partial_zV_{\mu})-\frac{Q^2}{g^2}V_{\mu} + 2 \kappa\varepsilon_{\mu\nu}\partial_zA_{\nu} &=& 0 \label{system},
\end{eqnarray}
where ${\mu,\nu}=0,1$, since we work in the gauge $A_z=V_z=0$. Again, we write down $V_{\mu}$ and $A_{\mu}$ fields through the mode and UV boundary functions
$V_{0\mu}$, $A_{0\mu}$:
\begin{equation}
V_{\mu}=V(Q,z)V_{0\mu}(Q),\,\,A_{\mu}=A(Q,z)A_{0\mu}(Q).
\end{equation}
To simplify further computations we work in the reference frame where $A_{00}=0$. It can be easily seen that the nontrivial solution can only be found in the case when $V_{01}=0$, $A_{01}\neq 0$ and $V_{00}\neq 0$. It means
\begin{equation}
V_{\mu}A^{\mu}\Big|_{z_0}=0
\label{perpendicular}
\end{equation}
that the gauge vectors $V$ and $A$ are perpendicular at the UV boundary. Of course, this condition does not hold in the bulk.
Eq.(\ref{perpendicular}) follows from the chiral algebra in $1+1$-dimensions where $j_{\mu}=\varepsilon_{\mu\nu}j^{\nu}$ for each
left and right components ($j_0^L=-j_1^L$, $j_0^R=j_1^R$), and the bulk relation $VA\sim L^2-R^2$. Using that only $A_{01}\neq 0$ and $V_{00}\neq 0$ 
are nonzero,   
we can write down explicitly system of differential equations for the mode functions $A(Q,z)$ and $V(Q,z)$:
\begin{eqnarray}
\partial_z(f^2(z)\partial_zA)-\frac{Q^2}{g^2}A + 2\kappa r\partial_zV &=& 0,\nonumber\\
\partial_z(f^2(z)\partial_zV)-\frac{Q^2}{g^2}V + 2\kappa\frac{1}{r}\partial_zA &=& 0,
\label{system2}
\end{eqnarray}
where we treat momentum $Q^2$, $\kappa$ and the ratio of sources $r=V_{00}/A_{01}$ as parameters. 
We solve this system numerically with boundary conditions:
\begin{eqnarray}
\textrm{IR brane:}\quad \partial_zV(Q,z)\Big|_{z=0}=0, \quad A(Q,0)=0, \\
\textrm{UV brane:} \quad V(Q,z_0)=1, \quad\quad\quad\, A(Q,z_0)=1.
\end{eqnarray}
and $z_0=\infty$. Further we justify that one can choose finite UV boundary values. 
To regulate the divergency we use $(3+\epsilon)$ dimensional regularization
with $(2+\epsilon)$ spacial dimensions where $\epsilon$ is small. The metric factors defined in eq.(\ref{metric-factor}) are 
\begin{equation}
f^2(z) =\frac{1}{g_3^2}\frac{g_{\mu\mu}^{\epsilon/2}}{\sqrt{g_{zz}}},\quad g^2(z) =g_3^2\frac{g_{\mu\mu}^{1-\epsilon/2}}{\sqrt{g_{zz}}},
\label{metric-factor-epsilon}
\end{equation}
where we omit factor of $2$ and include it in redefining the other constant, and the squared of coupling has dimension of mass
$[g_3^2]=m$ in the $3$-dimensional theory, for the metric factors $[f^2]=[1/g^2]=L$ with $L$ denotes the dimension of length.  
In $(3+\epsilon)$ dimensions the intergral defining the pion constant $f_{\pi}$ 
and the susceptibility $\chi$ (which are both dimensionless in $3$D)
\begin{equation}
\int_{-z_0}^{z_0}\frac{dz}{2f^2(z)} 
\end{equation}
becomes convergent (it will be clear in a concrete model). Let us estimate the Wronskian in the IR and see how the dimensional regularization works in this case. Using that $\partial_zV=0$ and $A=0$ around $z=0$, the first equation in the system (\ref{system2}) reduces to
\begin{equation}
\partial_z(f^2\partial_z A)=0,
\end{equation}
with solution given by
\begin{equation}
A(Q,z)=C\int_0^z\frac{dz}{f^2(z)},\quad C= \left(\int_0^{z_0}\frac{dz}{f^2(z)}\right)^{-1}
\end{equation}
Using the IR boundary conditions in the second equation of (\ref{system2}), we have
\begin{equation}
-\frac{Q^2}{g^2}V+2\kappa\frac{1}{r}\partial_zA=0,
\end{equation}
Substituting solution for $A$, we find
\begin{equation}
V(Q,z)=\frac{2\kappa C}{rQ^2}\frac{g^2(z)}{f^2(z)}.
\end{equation}
Using these solutions, we get for the Wronskian around $z=0$
\begin{equation}
W(Q^2,z)=f^2(VA^{\prime}-AV^{\prime})\rightarrow f^2 VA^{\prime} = \frac{2\kappa C^2}{r}\frac{1}{\Lambda^2Q^2}\frac{g^2(z)}{f^2(z)},
\label{wronskian-epsilon}
\end{equation}
where the $z$ dependence is given by the metric factors eq.(\ref{metric-factor-epsilon})
\begin{equation}
\frac{g^2(z)}{f^2(z)}=g_3^4 g_{\mu\mu}^{1-\epsilon}(z),
\end{equation} 
$\Lambda$ is introduced to make $z$ dimensionless (see further).
We plot the Wronskian around $z=0$ eq.(\ref{wronskian-epsilon}) in the "cosh" model with $g^2/f^2\sim (\cosh(z))^{2(1-\epsilon)}$
using different $\epsilon$ Fig.(\ref{wronskian}). Decrease in $\epsilon$ leads to a flatter dependence for $W$. It is a desirable result. 


In what follows we consider the "cosh" and Sakai-Sugimoto models \cite{erlich-katz-son-stephanov,son-stephanov}, 
and perform there numerical calculations of system (\ref{system2}).  
Using eq.(\ref{metric-factor-epsilon}), we have for the "cosh" model in $(3+\epsilon)$ dimensions
\begin{eqnarray}
f(z)=\Lambda\frac{1}{g_3}(\cosh(z))^{\delta},\\ 
g(z)=g_3(\cosh(z))^{1-\delta}
\end{eqnarray}
where $\delta=\epsilon/2$  ($\delta=0$ corresponds to $3$D and $\delta=1$ to $5$D), 
and we add the energy scale $\Lambda$ to make the radial coordinate $z$ dimensionless. 

We perform numerical calculations using "cosh" metric factors. 
We find diverging solutions in the $3$D 
are regulated, i.e.
become converging in the $(3+\epsilon)$D 
Wronskian is a constant for the Maxwell case. 



\begin{figure}[!ht]
\includegraphics[width=0.45\textwidth]{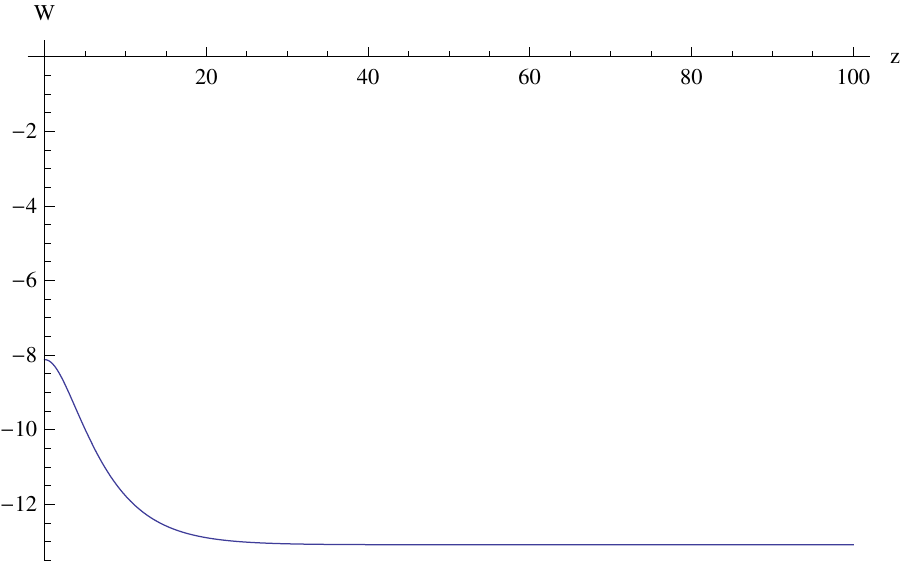}
\includegraphics[width=0.45\textwidth]{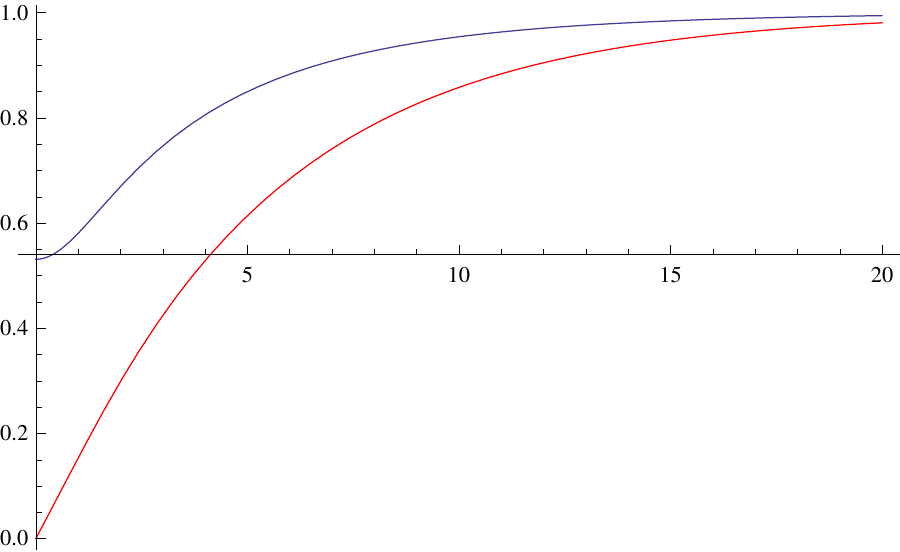}
\caption{"Cosh" model in $(3+\epsilon)$D. The Wronskian $W(Q,z)$ (left panel) and gauge functions $V(Q,z)$, $A(Q,z)$ (right panel) in the Maxwell-Chern-Simons theory.
Parameters are $Q=5$, $\Lambda=10$, $2\kappa=1$, $r=10$, $\delta=0.1$.}
\label{cosh4}
\end{figure}

Using the "cosh" metric factors, 
we add the Chern-Simons term in $(3+\epsilon)$D Fig.(\ref{cosh4}). We find that due
to the dimensional regularization the Wronskian developes a platou starting from some $z$. Also solutions for the gauge functions
converge to a finite value in the UV asymptotics. Solutions for $V(Q,z)$ and $A(Q,z)$ do not change much when the Chern-Simons term is included 
(with Chern-Simons the difference between solutions becomes slightly larger in the IR). Increasing $\kappa$  practically does not change the transition point
at which $W(Q,z)$ tends to a platou. 
We also don't see any crucial difference for the cases
$Q>\Lambda$ and $Q<\Lambda$.

With "cosh" metric factors, decreasing $\epsilon$, we find that the solutions $V(Q,z)$ and $A(Q,z)$ remain regular,
which produce $z$-independent Wronskian starting from some $z$. We observe numerically that the limit of small $\epsilon$ exists
with regular solutions. Diverging solutions appear exactly in $3$D. We suggest that the logarithmic divergence is an artefact of
$(2+1)$ dimensional theory and it can be regulated by the dimensional regularization.   


We also examin numerical solutions in Sakai-Sugimoto model. Solutions in this model express similar behavior, although we found "cosh" model is more suitable for numerical investigation. 

From eq.(\ref{metric-factor-epsilon}), we have for the Sakai-Sugimoto model in $(3+\epsilon)$ dimensions
\begin{eqnarray}
f(z)=\Lambda\frac{1}{g_3}(1+z^2)^{1/6+\delta},\\ 
g(z)=g_3(1+z^2)^{1/2-\delta}
\end{eqnarray}
where $\delta=\epsilon/6$ ($\delta=0$ is $3$D and $\delta=1/3$ is $5$D), and $\Lambda$ is added to make $z$ dimensionless.

Using Sakai-Sugimoto metric factors, we find solutions in the pure Maxwell theory. 
We see that the dimensionally regulated solutions
converge to a finite value in the UV. 

\begin{figure}[!ht]
\includegraphics[width=0.45\textwidth]{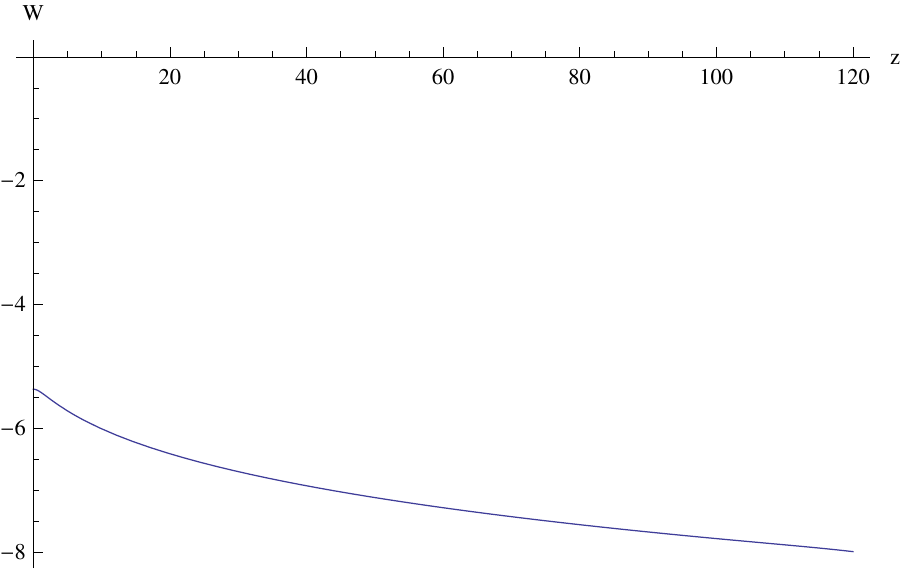}
\includegraphics[width=0.45\textwidth]{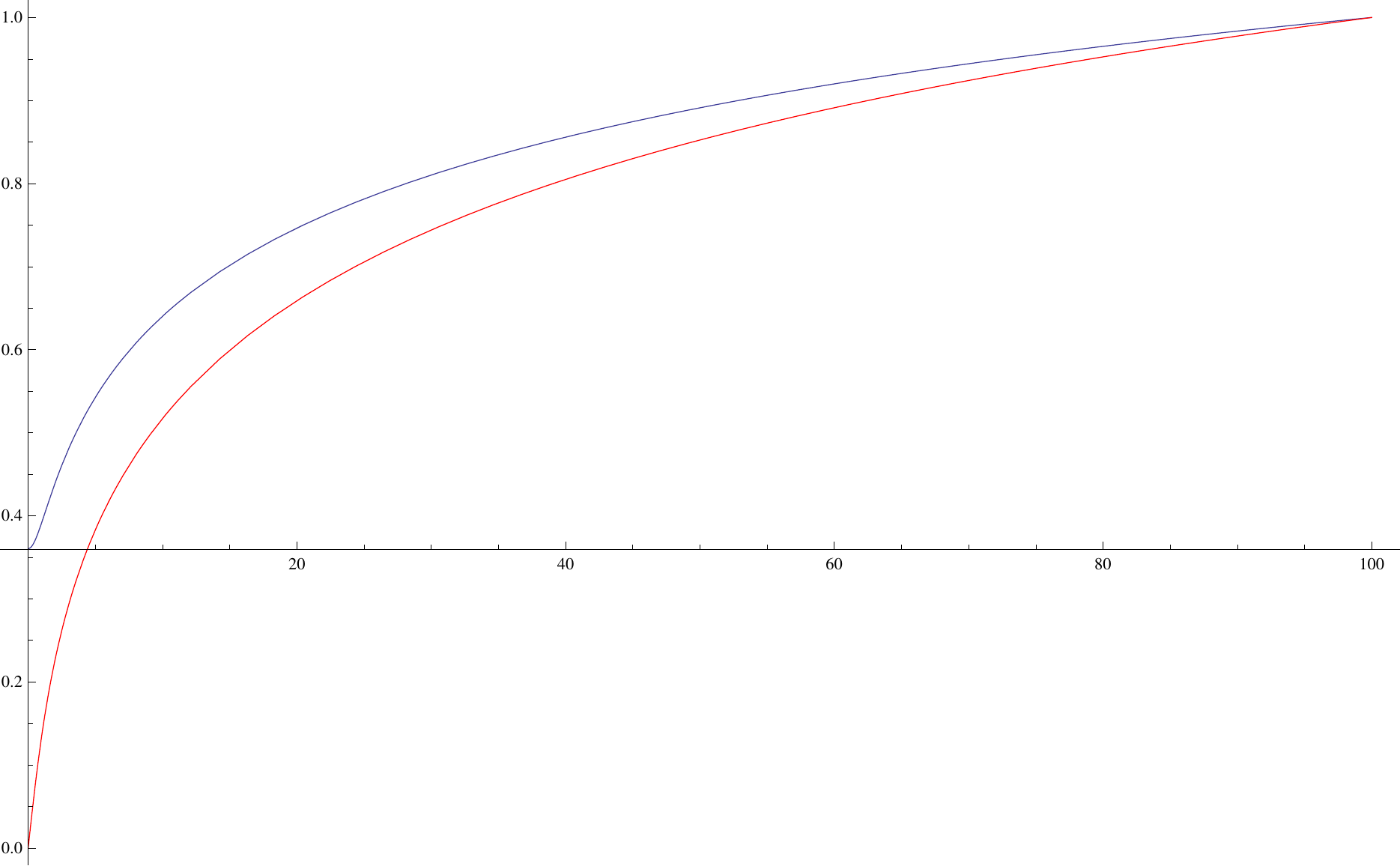}
\caption{Sakai-Sugimoto model in $(3+\epsilon)$D. The Wronskian $W(Q,z)$ (left panel) and gauge functions $V(Q,z)$, $A(Q,z)$ (right panel) in the Maxwell-Chern-Simons theory.
Parameters are $Q=5$, $\Lambda=10$, $2\kappa=1$, $r=5$, $\delta=0.1$.}
\label{ss3}
\end{figure}

With Sakai-Sugimoto metric factors, Wronskian (left panel) and solutions (right panel) in the Maxwell-Chern-Simons theory are displayed 
in Fig.(\ref{ss3}) in $(3+\epsilon)$D. The dimensionally regulated case in Fig.(\ref{ss3}) show
that the Wronskian tends to a platou and solutions are regular in the UV. Decreasing $\epsilon$ we find that this trend remains,
that suggests that the limit $\epsilon=0$ exists. 


Our numerical data justify the assumption that the Wronskian is independent of the radial coordinate for $z\gg1$. We find that adding the Chern-Simons
term does not solve the problem of logarithmically diverging solutions. We used the dimensional regularization in $(3+\epsilon)$D 
with small $\epsilon$ in order to regulate the gauge functions $V(Q,z)$ and $A(Q,z)$ which produce constant behavior for the Wronskian $W(Q)$.
It also justifies the use of the finite UV boundary conditions for $V(Q,z)$ and $A(Q,z)$.
    

\end{document}